\documentclass[reprint,amsmath,amssymb,aip,apl]{revtex4-2}

\usepackage[english]{babel}
\usepackage{graphicx}
\usepackage{float}
\usepackage{multirow}
\usepackage{hyperref}
\usepackage{lipsum}
\usepackage{csquotes}
\usepackage{rotating}
\usepackage{xspace}
\usepackage[table,xcdraw]{xcolor}
\usepackage{textgreek}

\hypersetup{colorlinks=true, allcolors=blue}

\begin{document}

\title{A Framework to Pinpoint Performance Bottlenecks in Emerging Solar Cells and Disordered Devices via Differential Machine Learning}

\author{Cai Williams}
\author{Chen Wang}
\author{Alexander Ehm}
\author{Dietrich R.~T.~Zahn}
\author{Maria Saladina}
\author{Carsten Deibel}
\affiliation{Institut für Physik, Technische Universität Chemnitz, 09126 Chemnitz, Germany}

\author{Roderick C. I. Mackenzie}
\affiliation{Department of Engineering, Durham University, Lower Mount Joy, South Road, Durham DH1 3LE, United Kingdom}

\date{\today}

\begin{abstract}
A key challenge in the development of materials for the next generation of solar cells, sensors and transistors is linking macroscopic device performance to underlying microscopic properties. For years, fabrication of devices has been faster than our ability to characterize them. This has led to a random walk of material development, with new materials being proposed faster than our understanding. We present two neural network-based methods for extracting key material parameters, including charge carrier mobility and trap state density, in optoelectronic devices such as solar cells. Our methods require solely measured light current--voltage curve and modest computational resources, making our approach applicable in even minimally equipped laboratories. Unlike traditional machine learning models, our methods place the final material values in a non-Gaussian likelihood distribution, allowing confidence assessment of each predicted parameter. We demonstrate these techniques using fresh PM6:Y12 and degraded PM6:BTP-eC9 organic solar cells.
\end{abstract}

\maketitle

\newcommand{\mcfigure}[4][0.65]{%
    \vspace{1em}
    \noindent\begin{minipage}{\linewidth}%
        \centering
        \includegraphics[width=#1\linewidth]{#2}
        \caption{#3}
        \label{#4}
    \end{minipage}
    \vspace{0.5em}
}

\newcommand*\rot{\rotatebox{90}}

\section{Introduction}
In recent years, there has been an explosive growth in the development of novel materials for optoelectronic devices, including organic polymers\cite{mazzolini2024pathways, simotko2025blending, chen2024dual}, small molecules\cite{li2024co}, and perovskites\cite{qiu2025over, li2024acceleration, pang2024reconfigurable}. Today, thin film analogues based on disordered materials exist for nearly all conventional inorganic devices, including organic LEDs \cite{Fan2025}, organic \cite{bae2025boosting, zahra2025trifluorophenyl,chen2025harnessing} solar cells, perovskite solar cells\cite{li2025boosting, wei2025surpassing, ding2024dopant}, transistors\cite{reginato2025improved, wang2022strain}, and sensors\cite{zheng2024highly, shin2025molecular}. Some of these technologies, most notably OLED displays, have seen significant commercial success. Nevertheless, progress across the field has been relatively slow. For instance, more than a decade after their modern rediscovery \cite{lee2012efficient}, perovskite solar cells have yet to achieve broad market adoption, and organic photovoltaics, despite their promise, remain largely confined to the lab. In order to see commercial success the learning rate of such systems must be increased.

If one takes the perovskite material system as an example, a cursory search on any on-line archive will reveal tens of thousands of articles published over the last 10 years, each reporting some improvement over the state-of-the-art. However, in most cases, characterization is limited to reporting of the overall device efficiencies (PCEs) based on current density--voltage (JV) curves, yielding the solar cell parameters short-circuit current density ($J_\mathrm{sc}$), fill factor (FF) and open-circuit voltage ($V_\mathrm{oc}$). Often, this is supplemented by basic measurements such as external quantum efficiency or absorption. However, researchers rarely employ and analyse techniques that truly clarify why changes in fabrication conditions or formulation/molecular structure affected device performance in terms of microscopic electrical parameters, such as trap densities, Urbach tail slopes, charge carrier mobilities, and recombination rates. Within organic photovoltaics, generally speaking, when a chemist changes the length of an aliphatic side chain on a novel acceptor molecule, it is common to report the change in efficiency, but not a detailed analysis as to what this did to the charge carrier dynamics and the core reasons for changes in efficiencies. This lack of closed loop feedback has turned device and material optimization into a random walk that spans decades. Consequently, it can take tens of years to turn a promising new material system into a commercially viable product; this is time we do not have in the context of the CO$_{2}$/energy crisis \cite{friedlingstein2023global}.

Techniques capable of probing complex physical processes, such as charge transport and recombination, remain underutilized due to their time-intensive nature and reliance on specialized equipment. To use these techniques, one often needs high speed pulsed lasers, fast electronics, and dedicated dark-room facilities. Consequently, access to these methods is often confined to a small number of well-resourced laboratories. Furthermore, expertise is compartmentalized across subdisciplines of materials science. For instance, techniques for analysing charge carrier dynamics are more likely to be found within condensed matter physics or physical chemistry groups, whereas the synthesis and processing of novel materials typically occur in synthetic chemistry laboratories. While cross-disciplinary collaboration is common in the field, this separation of skills and infrastructure makes it impractical to apply comprehensive characterization to every device that is fabricated. The rate at which new materials and devices are fabricated continues to accelerate outpacing the ability to meaningfully characterize them. Devices that underperform are frequently discarded without any detailed analysis, leaving the reasons for failure unknown.

One way to address this challenge is to apply modern machine learning (ML) techniques \cite{kingma2014adam, he2016deep, elfwing2018sigmoid,cao2025molecular}, with the aim of extracting key device parameters that are difficult to extract, such as mobilities and trap densities, from rudimentary measurement techniques. In this work, we combine state-of-the-art ML techniques based on residual blocks\cite{he2016deep} which are used in most modern Large Language Models (LLMs)\cite{achiam2023gpt,team2023gemini,grattafiori2024llama}, with synthetic data sets and experimental results. We demonstrate a ML framework that is capable of extracting key material parameters from single standard light JV measurement and, most importantly, places reasonable bounds on the accuracy of the results, without the need for frequency-domain characterisation.

It is the overarching aim of this paper to develop a method that can extract as much information as possible from a functioning device using the simplest possible measurements. We want to enable any laboratory that fabricates devices to be able to gain detailed information on their devices quickly and at low cost. To achieve this, we will focus on small datasets derived from minimal, low-complexity simulations and experiments. We prioritize steady-state measurements and simulations, deliberately avoiding frequency and time-domain techniques due to their higher experimental and computational complexity.

\begin{figure}
    \centering
    \includegraphics[width=0.8\linewidth]{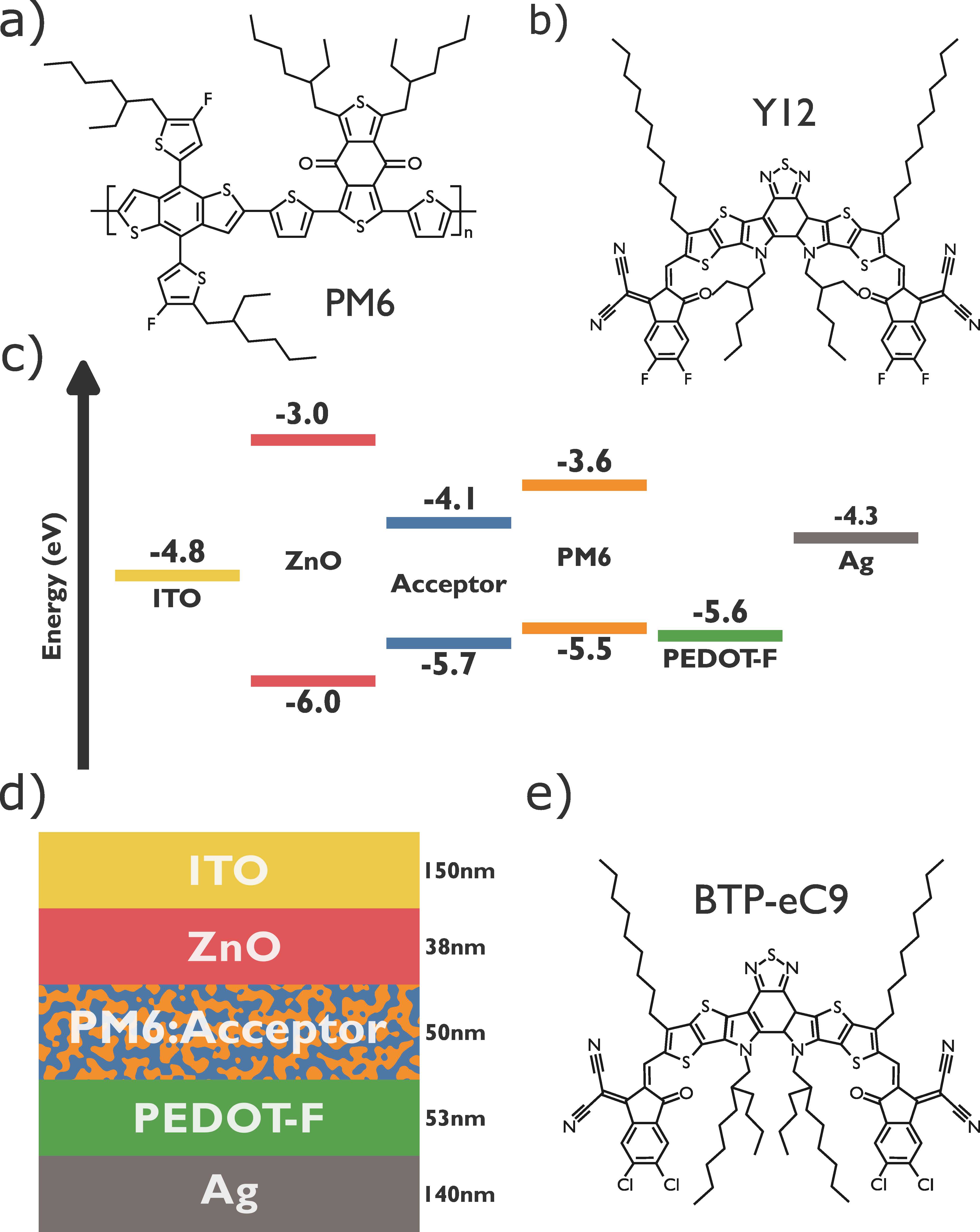}
    \caption{a) The molecular structure of organic donor PM6; b) The molecular structure of organic acceptor Y12; c) The energy level alignment diagram referenced to vacuum of both devices; d) The layer structure and thicknesses of the device within simulation; e) The molecular structure of organic acceptor BTP-eC9.}
    \label{fig:device}
\end{figure}

ML models are often criticized for producing predictions without clear justification, therefore incorporating a measure of confidence is essential. The method presented will quantify uncertainty in the extracted values. We introduce two key innovations: First, instead of predicting absolute values of the parameter of interest, the model is trained to learn the difference between a well-characterized reference device and a device under test. Second, this relative prediction enables a fully probabilistic output, where material parameters are represented as likelihood distributions. The form of the distribution is not assumed (unlike Bayesian based methods). We validate our approach on a series of PM6:BTP-eC9 and PM6:Y12 organic solar cells, including both freshly fabricated and aged devices, demonstrating that the method captures performance evolution over time. The acceptor molecules chosen represent the state-of-the-art for both chlorinated and fluorinated organic acceptor molecules. While this study focuses on organic photovoltaics, the approach is broadly applicable to a wide range of semiconductor systems and other classes of devices, offering a simple and accessible pathway toward interpretable, quantitative characterization.

\section{Experimental}
Devices were fabricated on pre-cleaned patterned ITO substrates with a 37~nm ZnO nanoparticle layer. PM6:BTP-eC9 and PM6:Y12 active layers (1:1.2 donor:acceptor ratio, 12~mg/mL in chloroform) were spin-coated at 45~$^\circ$C and annealed at 100~$^\circ$C, yielding 50--60~nm thick films. An ethanol-based PEDOT:F layer\cite{jiang2022alcohol} was deposited on top, followed by thermal evaporation of 150~nm thick Ag electrode to define a 0.04~cm$^2$ active area. All fabrication steps beyond ZnO deposition were conducted in a nitrogen glove box. See the supplementary information for further detail. The overall device structure was ITO/ZnO/active layer/PEDOT:F/Ag. This devices structure, and chemical structure of the donor and acceptors used can be seen in figure \ref{fig:device}.

\begin{figure}
    \centering
    \includegraphics[width=0.8\linewidth]{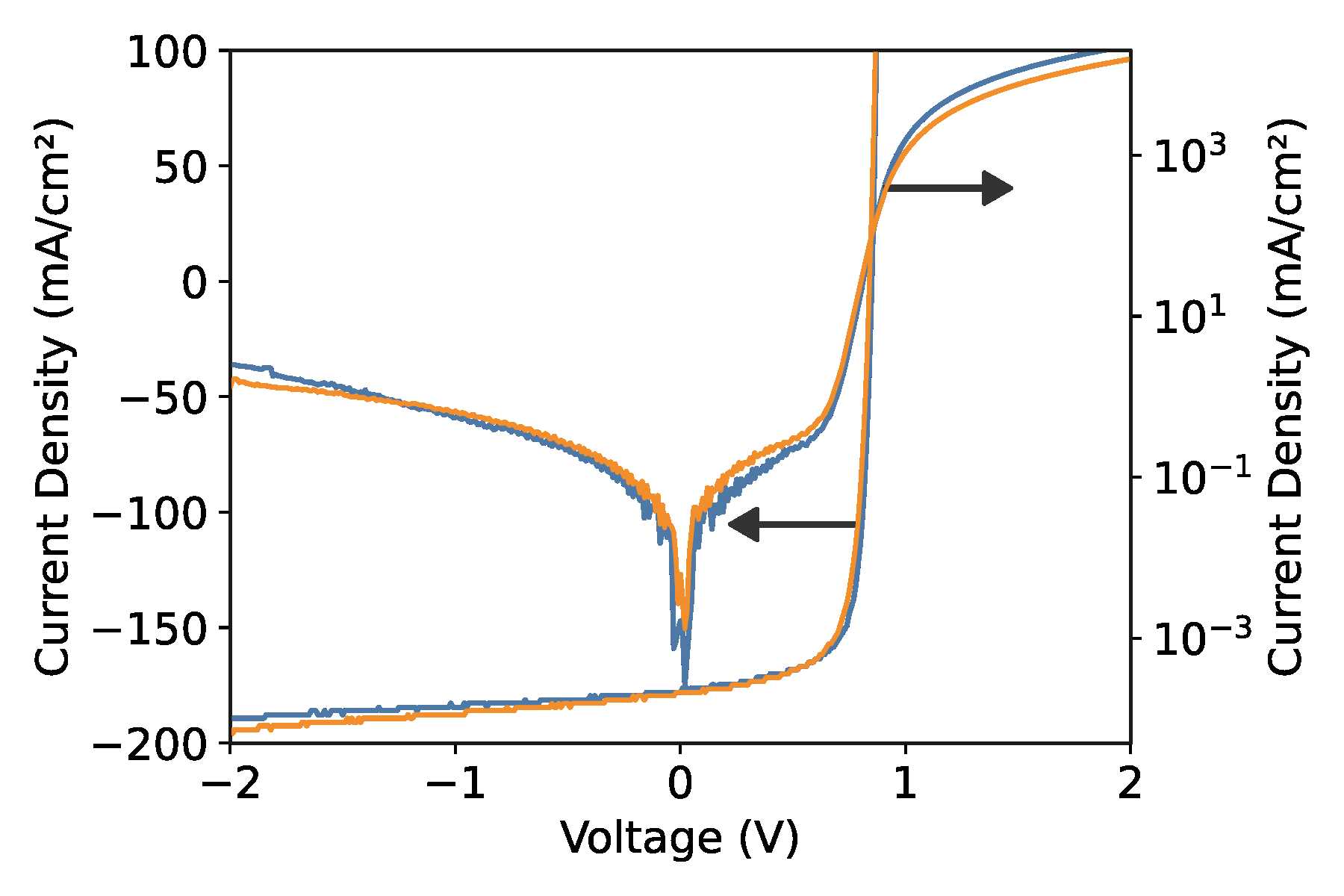}
    \caption{Light and dark JV curves for the PM6:Y12 (blue), and the PM6:BTP-eC9 (orange) device. The Light JV curves are indicated to the left, and dark JV curves indicated to the right.}
    \label{fig:exp_jv}
\end{figure}

For prediction, JV measurements were performed under AM1.5 illumination from a Wavelabs LS-2 solar simulator or Omicron A350 laser (515 nm), with experiments conducted in nitrogen atmosphere. For validation, Intensity-modulated photovoltage (IMVS) and photocurrent (IMPS) spectroscopy used modulated laser illumination and a Zurich MFLI lock-in amplifier, maintaining small-signal excitation with 10\% modulation depth. Space charge limited current (SCLC) devices were fabricated and mobility was extracted from hole- and electron-only devices using the Mott-Gurney law, with tailored layer stacks for selective carrier injection. Full details of the characterization techniques can be found in the supplementary information. It should be noted that IMPS, IMVS and SCLC are all relatively complex measurement not available in all labs, they are therefore only used in this paper for validation of the ML method and are not required to apply the ML method presented later.

Figure~\ref{fig:exp_jv} shows the light and dark JV curves for the PM6:Y12 and PM6:BTP-eC9 device. The PM6:Y12 and PM6:BTP-eC9 devices had a PCE of 11.0\% and 10.7\% respectively, this difference is due to the slightly higher $V_\mathrm{oc}$ of the PM6:Y12 device. The PM6:Y12 device was further characterised using IMPS.

\begin{figure}
    \centering
    \includegraphics[width=0.8\linewidth]{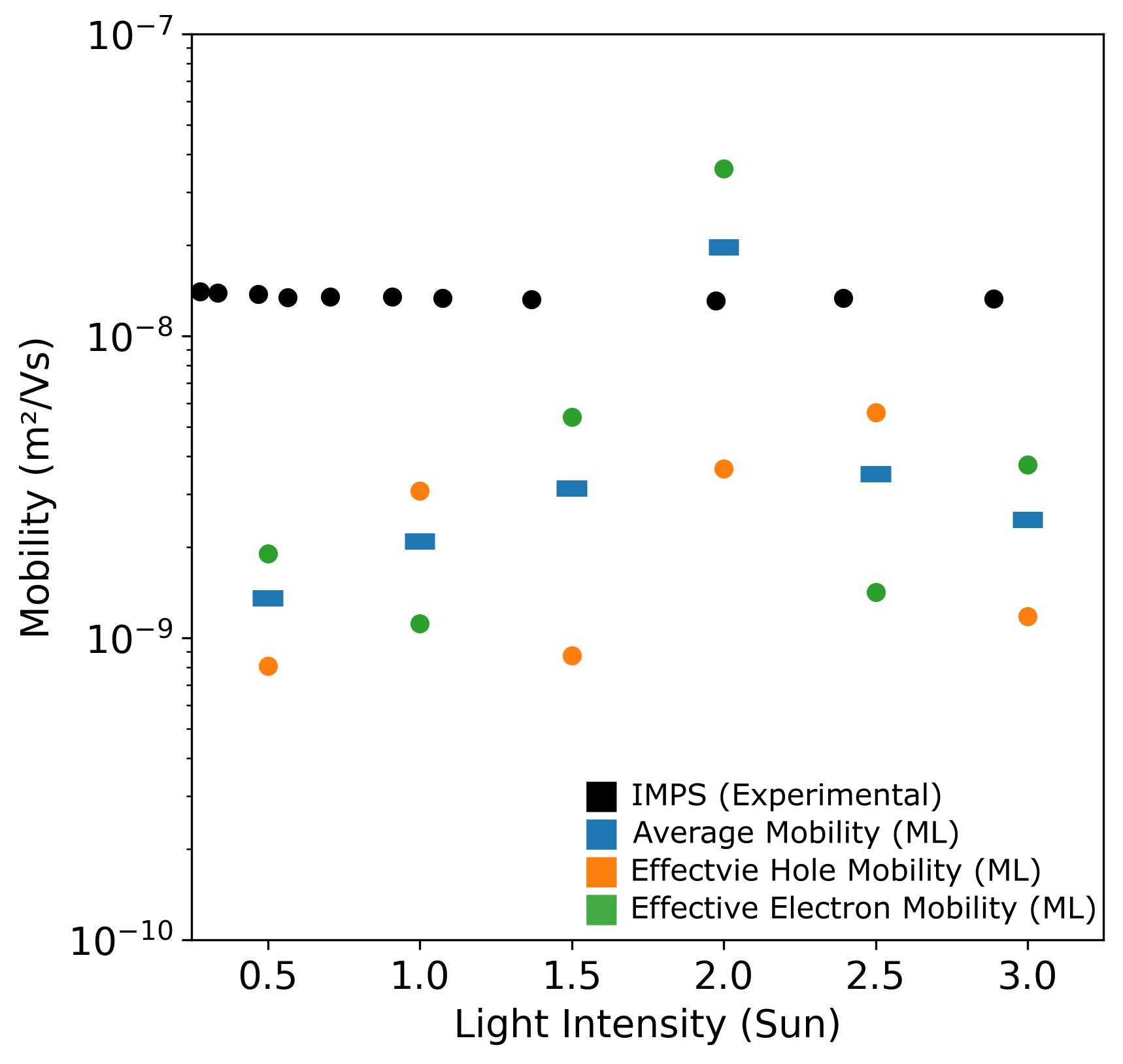}
    \caption{Experimental and predicted charge carrier mobilities. The black dots represent the values measured by IMPS, while other dots show ML-predicted mobilities for electrons, holes and their mean. The predicted mobilities are within an order of magnitude of the experimental values and remain largely independent of light intensity. The model struggles to differentiate between electron and hole mobilities, likely due to their close values. Note that the predicted values represent effective mobilities, averaged over both free and trapped charge carriers.}
    \label{fig:imps}
\end{figure}

\section{Dataset generation}
Modern neural networks require a significant amount of training data to produce reliable predictions, for most tasks they need to see between 10,000 and 100,000 examples. In practical terms, if one were trying to differentiate between cat and dog pictures, one would need a set of images already labelled by a human with the words cat or dog to train the network. In our case, producing an experimental data set on such a magnitude would be very challenging, as we would need to perform complex experiments on a large number of devices. Furthermore, many experimental techniques used to measure microscopic device quantities, such as trap density and mobility, often struggle to produce accurate values due to the variation of these parameters with charge carrier density \cite{hussner2024physical, kirchartz2025challenges}. Thus, were we to attempt to generate such a data set based on experimental data, we would risk learning incorrect values.  For these two reasons we used a synthetic data set generated by our drift--diffusion model that includes a distribution of trap states. It should be highlighted here that, because we are modelling disordered semiconductors, it is essential to use a model that includes carrier transport; carrier trapping, and recombination through trap states\cite{shockley1952statistics}. If one were to attempt to use a model without trap states, one would risk getting the carrier density in the device wrong, and thus the recombination rates/mobility values. This is discussed at length in the supplementary information.

To generate the data set, we first built the device structure of interest in our drift--diffusion simulation OghmaNano\cite{gao2015engineering,mackenzie2016loss}. We inputted the device geometry and also imported experimental complex refractive index ($n,k$) data for each layer (listed and explained in SI Section~\ref{si:sec:implemented_networks}). Using this base device simulation, a set of 1,000--10,000 child simulations with an identical structure were made. For each of these devices the electrical parameters such as free carrier mobilities, trap densities, Urbach tail slopes, capture cross sections, shunt and series resistances were randomly varied within defined bounds. A log scale was used where appropriate. Each of these simulations was then run to produce a light JV curve. This left us with a data set of light JV curves, and the exact electrical parameters used to produce them. Around 20\% of this data was held back for validation. We also removed devices with an efficiency below 8\% percent as devices of low performance otherwise become dominant within the datasets increasing the difficulty of training. This process could be completed in a couple of hours on a modern PC.

\section{A simple neural network for point prediction}

\begin{figure}
    \centering
    \includegraphics[width=0.8\columnwidth]{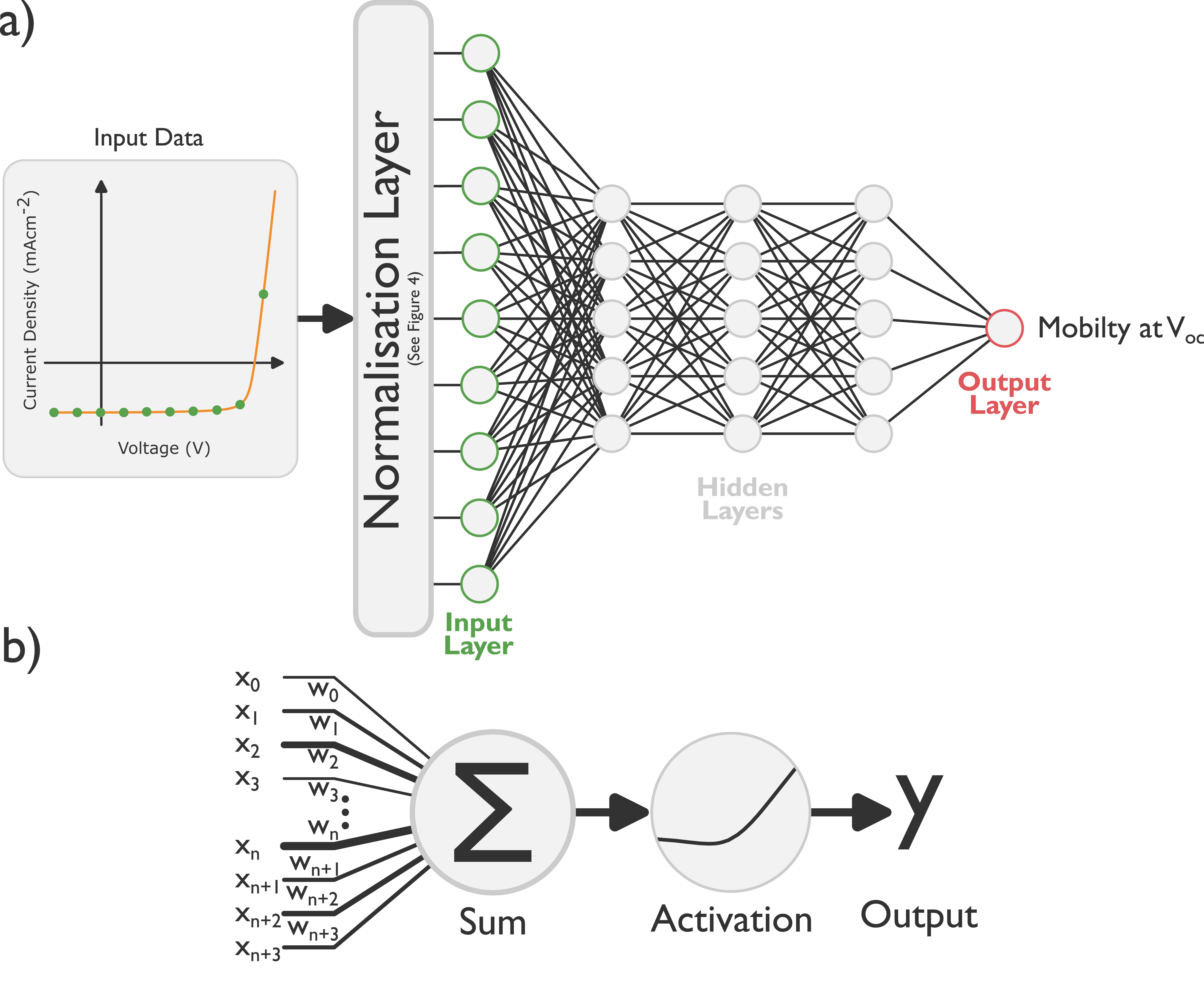}
    \caption{a) A neural network consisting of an input layer, hidden layers and an output layer. The JV curve of interest is placed on the input by discretizing it at the points indicated by the green dots. The network then predicts the mobility that was used to generate the JV curve. b) An expanded view of a single neuron, where inputs are weighted, summed, and passed through an activation function that determines the neuron's output.}
    \label{fig:point-nn}
\end{figure}

\begin{figure*}
    \centering
    \includegraphics[width=0.7\textwidth]{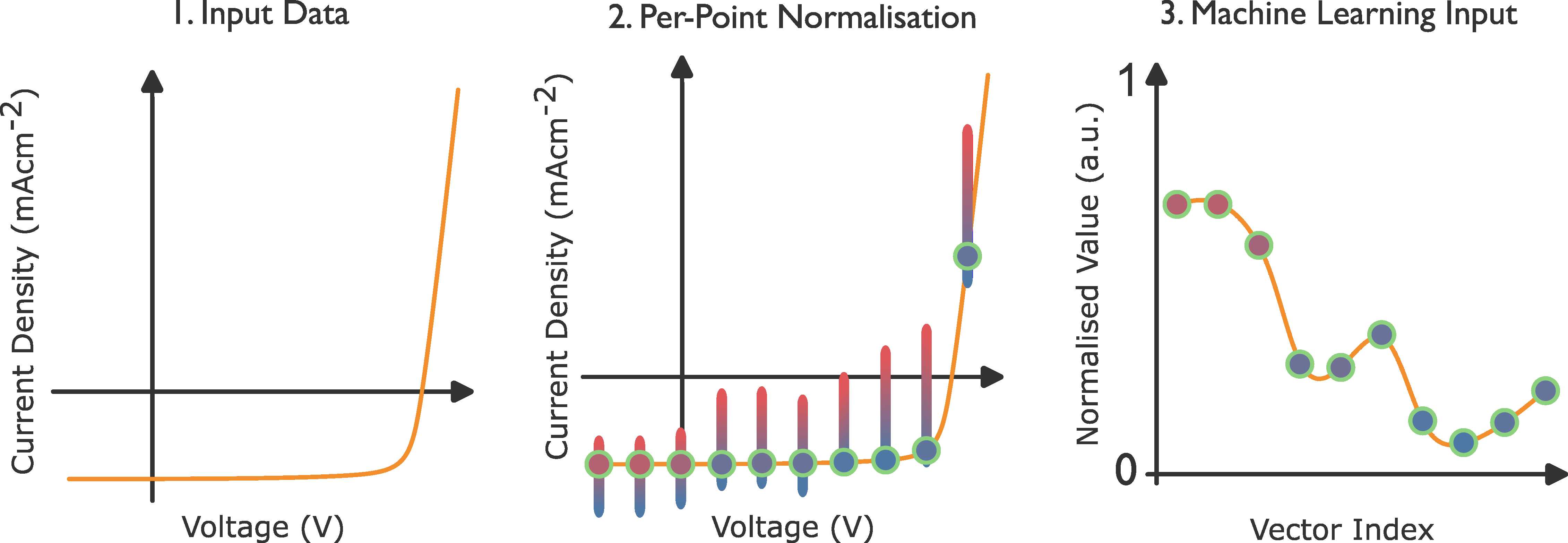}
    \caption{Normalisation process of input data: a) Experimental current--voltage data; b) Green dots represent voltages at which the current values are extracted, with red/blue bars showing the maximum/minimum range of the values across the entire dataset at each corresponding voltage; c) The normalized input to the neural network. Note the points are normalized relative to all data points in the dataset at a given voltage, not relative to the rest of the curve.}
    \label{fig:nrom}
\end{figure*}

We first use a deep neural network in the most simple and classical way that we can \cite{majeed2020using, hussner2024machine}, to baseline more advanced methods described later. This can be seen in Figure~\ref{fig:point-nn}a, with a JV curve on the left hand side of the image and a predicted value on the output. The neural network, once trained, acts as a transform between the input data and the predicted output. In this case, the network is extracting mobility from a JV curve, but the parameter could just as well be any material property, such as trap density. The actual size of the network is much larger than depicted in the figure and is detailed in the SI. Each node of the network is made up of an artificial neuron, schematically shown in Figure~\ref{fig:point-nn}b. Each neuron takes a series of inputs, first multiplying them by the weight $w_n$. The sum of these values is then applied to activation function allowing the output of the neuron to be non-linear, and when combined together universal function approximators. In this work, we use the Sigmoid Linear Unit (SiLU) activation function \cite{elfwing2018sigmoid}, as it is widely used in large models such as LLMs. Initially, the network is assigned a set of random weights. It is then trained by presenting known input data (e.g., a JV curve) and comparing the network’s output (e.g., mobility) to the known target value used to generate that input. The difference between the predicted and true values defines an error function, which is used to update the weights through backpropagation \cite{rumelhart1986learning, kingma2014adam}. Repeating this process thousands of times over the training dataset yields a network capable of predicting the desired material parameter.

Key to having a well performing network that takes note of all parts of the data is to correctly normalize the input data. If one considers the current density of a dark JV curve (e.g., Figure~\ref{fig:exp_jv}), the values at 0.1~V are up to 4 orders of magnitude smaller than those at around 1.5~V. Thus, feeding such a curve into a neural network would cause the high-current values to dominate, while lower-current values would be effectively ignored. To mitigate this imbalance, the data is normalized, as shown in Figure~\ref{fig:nrom}. 

The input data can be seen in Figure~\ref{fig:nrom}a, and in Figure~\ref{fig:nrom}b the green dots highlight specific voltages at which the current density values are sampled from the JV curve, while red/blue bars indicate the maximum and minimum range across the whole dataset at these voltages. Figure~\ref{fig:nrom}c presents the final normalized input to the neural network. 
Note that normalization is performed relative to all data points in the dataset at a given voltage; not relative to the rest of the curve. This ensures that all input values fall between 0.0 and 1.0. The same normalization approach is applied to the output data, and logarithmic scaling is used where appropriate. 

\begin{figure*}
    \centering
    \includegraphics[width=0.8\linewidth]{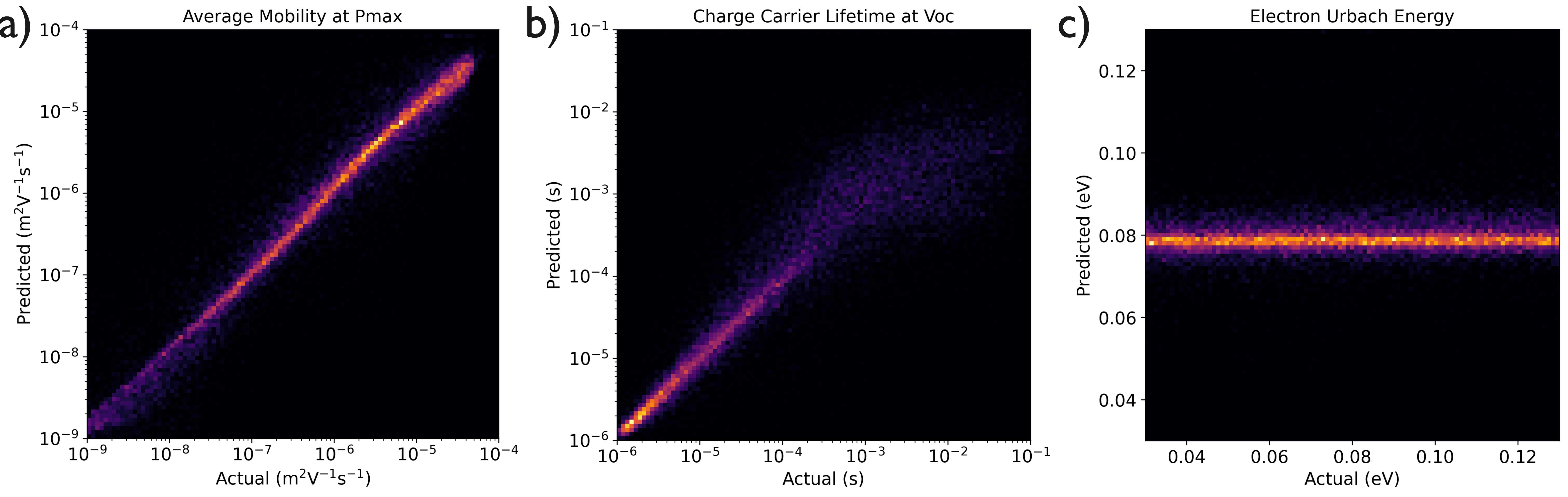}
    \caption{Confusion matrices for a) average carrier mobility at the maximum power point $\mu_\mathrm{pmax}$, b) charge carrier lifetime at open-circuit voltage $\tau_\mathrm{voc}$, and c) electron Urbach energy $E_U^e$. Average carrier mobility at the maximum power point is straightforward to predict. Charge carrier lifetime at $V_\mathrm{oc}$ is also reliably predicted, provided sufficient training data is available. However, Urbach tail slope is very difficult for a simple feed forward neural network to extract, simply because there is not much information about the Urbach tail slope in a single light JV curve.}
    \label{fig:dif-confusion}
\end{figure*}

Once trained, the network can be tested on the 20\% of the data set that was not used for training. Figure~\ref{fig:dif-confusion} shows the results from these tests plotted as so called confusion matrices. Here we show confusion matrices for the mobility at the maximum power point $\mu_\mathrm{pmax}$, charge carrier lifetime at open-circuit voltage $\tau_\mathrm{voc}$, and electron Urbach energy $E_U^e$. To generate these graphs, the normalized values for each JV curve are placed on the inputs of the network, the true value of mobility (or other parameter) used to generate the curve are placed on the x-axis of the plot, and the predicted value from the network is then placed on the y-axis. For a perfect prediction, the plot should be a straight line. The fewer off diagonal elements that exist, the better the ML system performs.

The three parameters presented here were chosen because they display very different learning properties (plots of other parameters are available in the SI). From Figure~\ref{fig:dif-confusion}a, it is evident that the model accurately predicts the average charge carrier mobility at the maximum power point ($\mu_\mathrm{pmax}$), as all points fall on or near the diagonal. The model also performs well in predicting $\tau^e_\mathrm{pmax}$ for values well below $1\times 10^{-3}$~s, but struggles above this threshold due to insufficient training data: very few devices in the training set have recombination lifetimes as long as 1~ms, which would correspond to exceptionally, and perhaps unrealistically, high-performance devices.  In contrast, the Urbach energy for electrons ($E_U^e$) cannot be predicted by the model at this stage, simply because there is not a lot of information about the tail slope embedded within the IV quadrant of the JV curve and what information exists in the curve is hard to extract. Later we discuss a more sensitive method, that shows more success.

Once the network was properly trained and tested, we applied it to the experimental JV curves from Figure~\ref{fig:exp_jv}, extracting electron and hole mobilities as a function of light intensity. The resulting average mobility values, shown in Figure~\ref{fig:imps}, are in close agreement with those obtained via IMPS. Notably, all values were inferred from a single light JV curve measured at 1~Sun. 

However, we cannot assign error bars to these values. The best we can do is look at the confusion matrix to assess how well the network predicts $\mu_\mathrm{Jsc}$ and thereby determine whether the results are trustworthy. A closer look at the graph reveals that the predicted electron and hole mobilities alternate, with electron mobility sometimes exceeding hole mobility and vice versa. This indicates that the model struggles to consistently distinguish between the two charge carrier species, likely because their mobilities are close in magnitude. Since the network only provides point predictions, no further information can be extracted at this stage. The issue of trust and uncertainty in these predictions is addressed in the following two sections.

\section{The differential objective method}

The method described in the previous section provides estimates of material parameters, and when combined with the confusion matrices, it offers a reasonable indication of how much confidence one can place in each prediction. However, challenges arise when an experimental input lies outside the distribution of the training data, for instance, an s-shaped JV curve \cite{wagenpfahl2010} may produce predictions with no meaningful confidence, as it deviates significantly from what the model has learned. Here, we introduce a novel approach that outputs a probability distribution representing the confidence in each predicted parameter. Importantly, unlike Bayesian methods, our technique does not require assuming a prior distribution for the calculation.

\begin{figure}
    \centering
    \includegraphics[width=0.8\columnwidth]{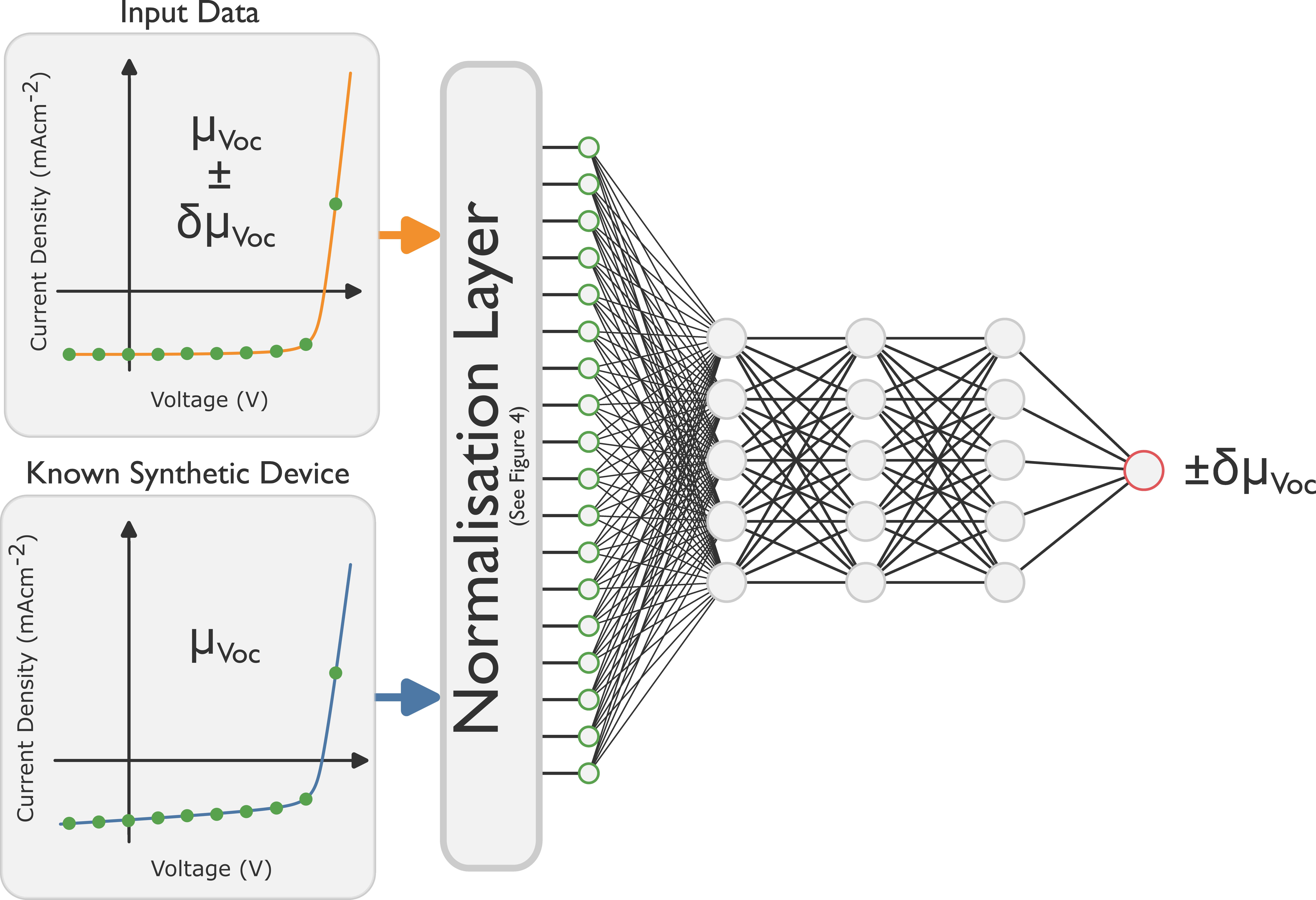}
    \caption{Application of the difference method: The network receives a 'known curve' with a known parameter (e.g., $\mu_\mathrm{voc}$) and an 'unknown curve'. Instead of predicting an absolute value, it estimates how much higher or lower the parameter value is in the unknown curve relative to the known one, with an output $\pm \delta \mu_\mathrm{voc}$. Repeating this over many known curves yields a distribution of parameter values predicted by the network.}
    \label{fig:dif-method}
\end{figure}

The method is depicted in Figure~\ref{fig:dif-method}. Instead of taking a single JV curve and asking the model to predict a single parameter, we present the network with two JV curves. One curve comes from the simulated dataset, where the parameters used to generate it are known ($\mu_\mathrm{voc}$ in the example shown in Figure~\ref{fig:dif-method}), and is referred to as the 'known curve'. The other curve, for which the generating parameters are unknown, is referred to as the 'unknown curve'. Rather than asking the model to predict a single value of mobility (or another parameter), we ask it to estimate how much higher or lower it believes the mobility in the 'unknown curve' is relative to the 'known curve', i.e., $\pm \delta \mu_\mathrm{voc}$. By adding $\pm \delta \mu_\mathrm{voc}$ to $\mu_\mathrm{voc}$, we obtain the estimated mobility for the 'unknown curve'. Repeating this process over many known curves allows us to obtain a distribution of predictions, and thus we can quantify how reliable the prediction is.

\begin{figure}[]
    \centering
    \includegraphics[width=0.8\columnwidth]{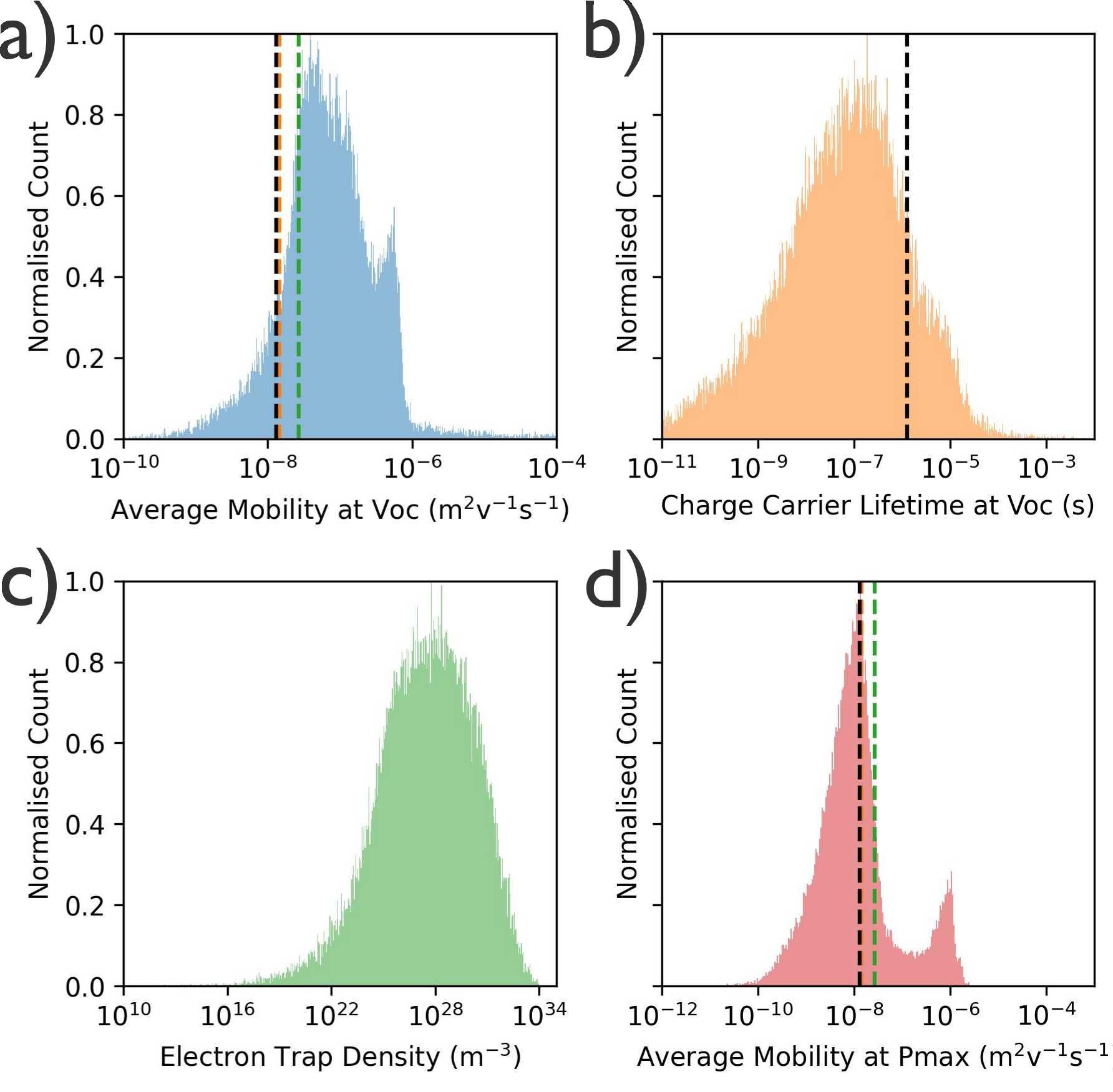}
    \caption{Difference network predictions for;
    a) Average Mobility at $V_\mathrm{oc}$;
    b) Charge carrier lifetime at $V_\mathrm{oc}$;
    c) Electron trap density;
    d) Average mobility at the maximum power point. 
    Some parameters can be predicted more accurately than others, each showing a unique, non-Gaussian distribution. In a and c, dashed lines indicate experimentally determined mobility values measurements: SCLC electron-only device (green), SCLC hole-only device (orange), and IMPS (black). The black line in b represents IMVS measured charge carrier recombination time. }
    \label{fig:dif-results}
\end{figure}

This is shown in Figure~\ref{fig:dif-results}, where we present histograms indicating where the model thinks the most likely material parameters lie for the experimental PM6:Y12 device. Four histograms are shown: average electron/hole mobility at $V_\mathrm{oc}$, charge carrier lifetime at $V_\mathrm{oc}$, electron trap density, and average electron/hole mobility at the maximum power point. For this device, the mobility at $V_\mathrm{oc}$ can be predicted with reasonable confidence within two orders of magnitude. Experimental SCLC mobility values are also included as vertical lines; orange for hole mobility, green for electron mobility, and black for the average. The distribution centres around $1\times 10 ^{-7}$~m$^2$~V$^{-1}$~s$^{-1}$, which compares well to the experimental SCLC values. As always, however it should be noted that experimental values of mobility in disordered materials are always difficult to accurately measure \cite{10.1063/1.4818267}. With that in mind, the predictions and measurements show very good agreement. In contrast to the charge carrier mobility, the distribution for charge carrier lifetime in Figure \ref{fig:dif-results}b is broader, but has a peak near $5\times 10 ^{-7}$~s. That value coincides with the lifetime measured by IMVS, which is $1.3\times 10^{-6}$~s under 1~Sun illumination. However, again it is worth noting that in disordered systems different measurement techniques often extract slightly different values for constants due to the dependence of most parameters on charge density.

Figure~\ref{fig:dif-results}c demonstrates that the electron trap density is, as expected, more difficult to extract as shown by the broad distribution generated. Nevertheless we can state that the value probably lies around $1 \times 10 ^{28}$~m$^{-3}$, which is the peak of the distribution. The average electron/hole mobility at the maximum power point is shown in Figure~\ref{fig:dif-results}d, and we can say with confidence that it is very likely to be $1\times 10 ^{-8}$ m$^2$ V$^{-1}$ s$^{-1}$ due to the sharp peak in the probability distribution. The SCLC lines are also plotted for comparison. It is interesting to note, when comparing Figure~\ref{fig:dif-results}a and Figure~\ref{fig:dif-results}d, that the distribution of mobilities in the latter is shifted to lower values than in the former. This is expected, as the carrier density is higher at $V_\mathrm{oc}$, resulting in a higher average charge carrier mobility.

The method described above has three distinct advantages: Firstly, one can visually assess where the value of interest is most likely to lie. Secondly, our method does not assume a prior distribution, unlike many approaches based on Bayesian statistics; for example, a Gaussian distribution would not capture the shapes shown in Figures~\ref{fig:dif-results}a and \ref{fig:dif-results}d. Thirdly, when it comes to training, it can dramatically reduce the required dataset size. In a dataset of 1,000 curves, we can train against all permutations of the curves, yielding an effective dataset size of 999,000. This not only decreases data generation time but also boosts the quality of the trained network.

\section{The differential objective method with residual blocks.}

\begin{figure}
    \centering
    \includegraphics[width=0.8\columnwidth]{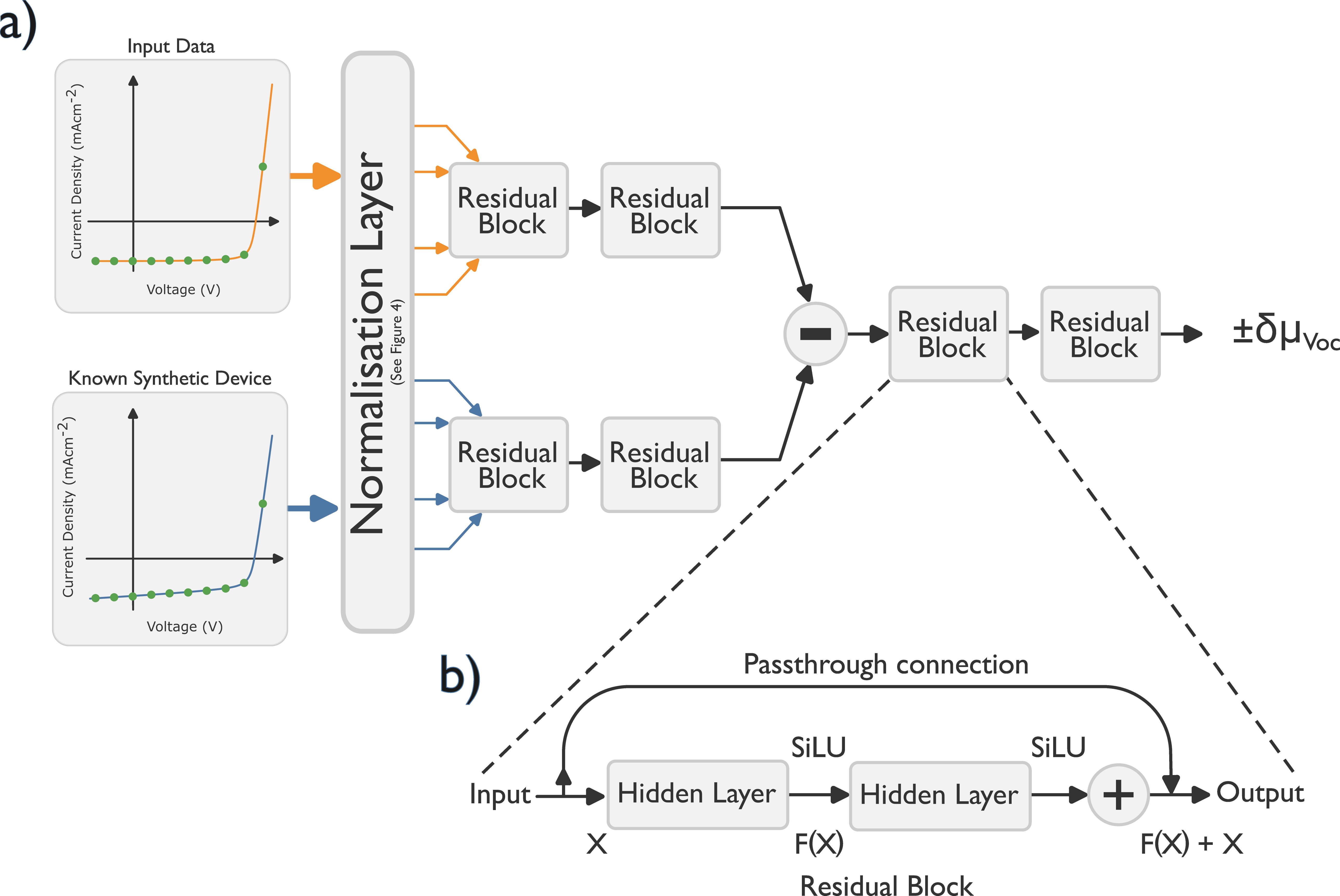}
    \caption{Implementation of the difference method with residual blocks. a) Schematic of the implemented network; the actual network includes additional residual blocks beyond those depicted. b) A residual block showing how the input is added to the output.}
    \label{fig:resid-network}
\end{figure}

In deep neural networks such as ours, training becomes more difficult as the network depth increases. This is because adjusting a weight in the centre of the network often has little impact on the quality of the prediction, making it hard to determine which way to tune a weight \cite{he2016deep,bengio1994learning}.  The overall effect of this to limit how deep a network can be before additional layers do not help with the task at hand. In the previous example we found that having more than four hidden layers did not add much to the predictive capabilities of the model. Another issue with deep networks is that neurons deeper in the network often lose sight of the input data and are only working on a transformed version of it. As a result, they may struggle to extract meaningful information. A solution to both these issues, proposed by He et al.,\cite{he2016deep} is the residual block, shown in Figure~\ref{fig:resid-network}b. The core idea of this block is that some fraction of the input data always bypasses the neurons and is routed directly toward the output. This allows deeper neurons to retain access to the information available at shallower layers of the network. This strategy is used to great effect by many modern LLMs \cite{achiam2023gpt, team2023gemini, grattafiori2024llama}. Figure~\ref{fig:resid-network}b presents the structure of our difference network with residual blocks included.  The residual block allows us to use a network that is much deeper. For comparison, in the pervious section our network was 4 layers deep with 226,049 trainable parameters, while the network in this section was 32 layers deep with 1,007,233 trainable parameters. It should be noted that this is still small in comparison with most LLMs, which typically have have 1-1.8 trillion weights.

\begin{figure*}
    \centering
    \includegraphics[width=0.8\linewidth]{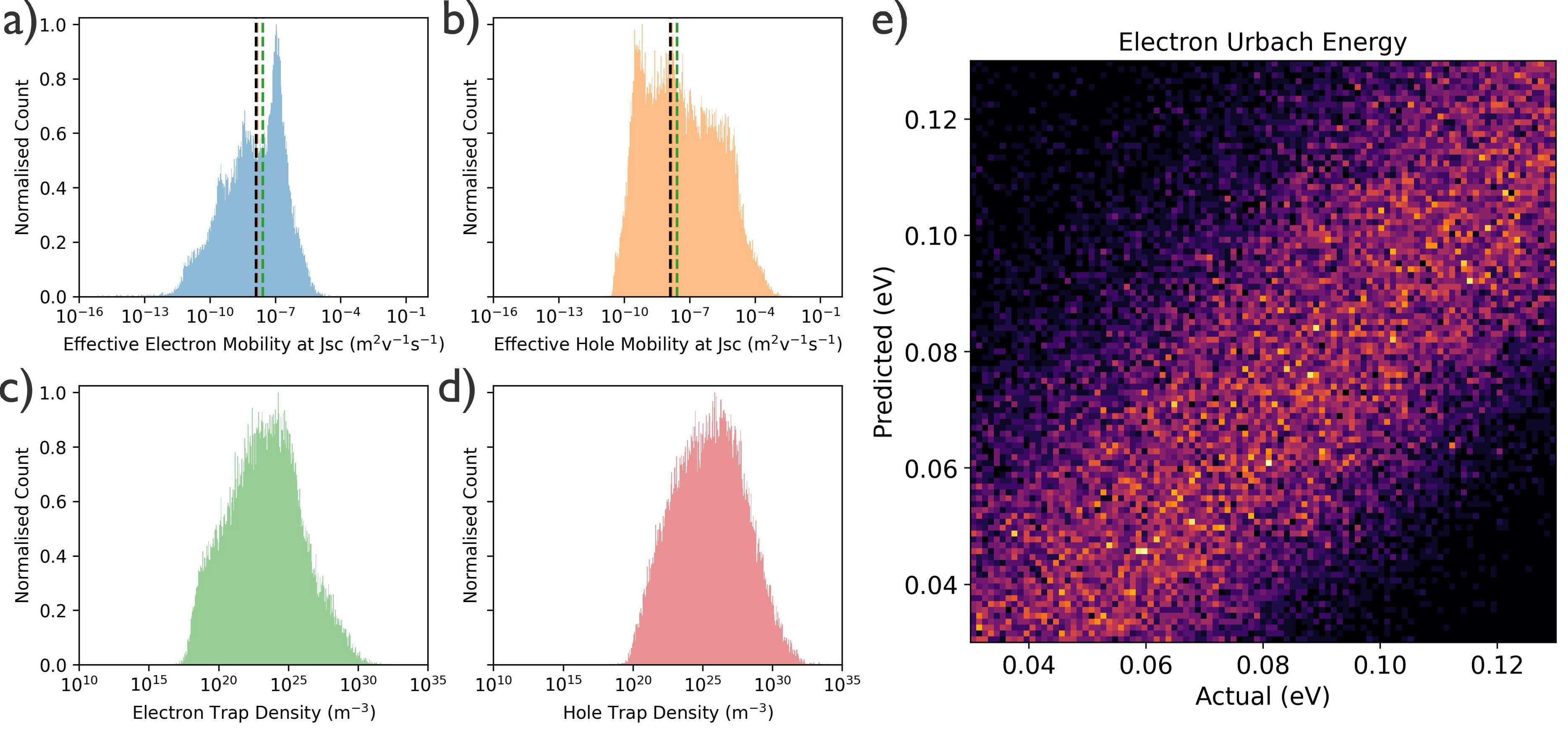}
    \caption{Residual network predictions for; 
    a) Effective electron mobility at $J_\mathrm{sc}$; 
    b) Effective hole mobility at $J_\mathrm{sc}$; 
    c) Electron trap density, and 
    d) Hole trap density. 
    In panels a, and b, dashed lines represent experimentally determined mobility values from SCLC measurements: electron-only device (green), hole-only device (orange), and IMPS mobility mean (black). Sub figure e) shows a confusion matrix for electron Urbach energy prediction.}
    \label{fig:resid-block-results}
\end{figure*}

Results from the residual block approach are shown in Figure~\ref{fig:resid-block-results}.
Overall we found that networks with the residual networks are better at extracting hard to learn parameters, such as Urbach energy, trap density and also energetically averaged charge carrier mobility but due to their larger size, longer training times were needed. Generally in terms of RMS error, residual block networks outperformed the networks in the previous section by a factor of between 2 and 1000, this can be seen in table \ref{tab:Net_Comp} in the SI.

Figure~\ref{fig:resid-block-results}a and Figure~\ref{fig:resid-block-results}b display the energetically averaged electron and hole mobility at $J_\mathrm{sc}$. The distribution of predictions for electron mobility is very sharp, indicating strong predictive performance, while the distribution for holes is more square-shaped, suggesting greater difficulty in predicting the value. This ties in well with the results in Figure~\ref{fig:imps}, where the model struggles to understand whether electron or hole mobility is higher. It is also worth noting that both distributions fall within roughly the same range, which is supported by previous work\cite{vollbrecht2023relationship}. As in the previous figures experimental lines are also shown. It should be emphasized that disentangling electron/hole mobility, rather than stating an average value is a difficult task even a well equipped lab.

In the previous section we showed that trap density (see Figure \ref{fig:dif-results}c) was hard to extract and the network predicted a broad distribution of $1 \times 10^{16} - 1 \times 10^{34}$~m$^{-3}$. This distribution has been recalculated in Figures~\ref{fig:resid-block-results}c and d for both electrons and holes. It can be seen the distributions have narrowed, with peaks at approximately $1 \times 10^{26}$~m$^{-3}$. These are reasonable parameters when considers that each molecule or polymer subunit capable of trapping charge might occupy a volume of $V=(1 \times 10^{-9})^{3}$~m$^{-3}$, and this would yield an expected trap density of around $1 \times 10^{27}$~m$^{-3}$ - were all sites efficiently connected to conducting pathways.

Figure~\ref{fig:resid-block-results}e, plots a confusion matrix of predicted and extracted Urbach energy. If one compares this to the previous confusion matrix shown in Figure~\ref{fig:dif-confusion}c, it can be seen that there is a considerable improvement, with the network being able to identify the $x=y$ trend. Although, the result is far from perfect, it should be taken in the context that in a single light JV curve there is very limited information about Urbach energy. Indeed to-date there are no reliable analytical methods, that have been proposed to extract Urbach energy from a single light JV curve.

\section{Application to degradation}

\begin{figure*}[t]
    \centering
    \includegraphics[width=0.8\linewidth]{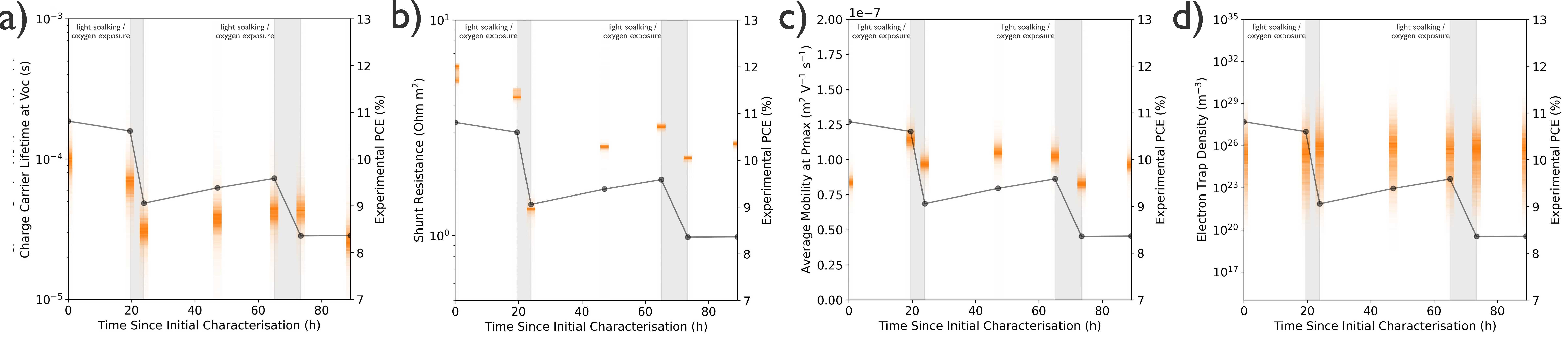}
    \caption{Residual network predictions upon degrading PM6:BTP-ec9 devices for;
    a) Charge carrier lifetime at $V_{oc}$;
    b) Shunt resistance;
    c) Average mobility at $P_\mathrm{max}$,
    and d) Electron trap density.
    In each figure the measured experimental PCE during degradation is shown as a reference for overall device condition.}
    \label{fig:deg}
\end{figure*}

We now apply the method developed above to a degradation study of PM6:BTP-eC9 device. Degradation studies provide a good testing ground for the method because the solar cell changes over time and there is no simple way to pause the process to perform complex charge carrier measurements. The experiments can also produce large volumes of data that are time-consuming to interpret using standard fitting techniques. We aged the solar cell in ambient air at $V_\mathrm{oc}$ under illumination from a 1000~W\,m$^{-2}$ Xenon Lamp, in accordance with the ISOS-L1 \cite{reese2011consensus} testing protocol. To increase confidence in the reported results and separate light soaking effects, due to the ZnO electrode \cite{kim2012light,gunther2023neglected}, from true degradation, the device was aged in stages. After fabrication, it was left in the glove box to stabilise for 20~h, followed by 4~hours of exposure to light and air. The device was then returned to the glove box for two days to allow light soaking effects to subside. This was followed by another 8~h of light exposure in air, and another rest period in the glove box. Periodic JV measurements were taken throughout the entire process.

If one examines the line in Figure~\ref{fig:deg}a representing PCE, the effect of this staggered degradation becomes apparent. Between the first two black dots, there is a gradual reduction in PCE, likely indicating device stabilization after fabrication due to residual solvent evaporation or molecular relaxation. During the first ageing step, a sharp drop in efficiency occurs, potentially due to oxygen reacting with donor polymers and moisture ingress.  After the device is returned to the glove box for 20~hrs, the efficiency increases by a few percent, followed by a rapid decline in PCE during the next ageing step.  This behaviour suggests that the device undergoes initial burn-in, partial recovery in an inert atmosphere, and then accelerated degradation upon further exposure.

The left-hand axis of Figure~\ref{fig:deg}a shows the predicted carrier lifetime, with the orange spots representing the probability distributions of the predicted values. It can be seen that the predicted recombination rates follow the trend of the PCE, mirroring degradation and recovery as it happens. Figures~\ref{fig:deg}b,c and d plot the shunt resistance, effective electron mobility at $P_\mathrm{max}$ and electron trap density. Shunt resistance and effective mobility follow the trends of PCE, while average mobility and as far as we can tell trap density do not change.

The reduction in shunt resistance is likely due to moisture-induced leakage pathways.  While, the decline in charge carrier lifetime probably reflects the formation of non-radiative recombination centres, such as oxidised species or defect states, either within the bulk or at critical interfaces. In contrast, the stability of charge carrier mobility suggests that long-range transport remains largely unaffected, potentially due to preserved molecular packing. The invariance in trap density further implies that deep trap states are not forming in significant numbers to be detected or the newly introduced traps are too shallow to be seen.

\section{Conclusions}

We present two practical methods for extracting key material parameters, such as charge carrier mobility and recombination constants, from experimental JV measurements. By employing a differential learning framework, our approach expresses material properties as probability distributions rather than point estimates, enabling direct quantification of uncertainty. Performance is enhanced through the use of residual blocks, which improve the accuracy and robustness of the deep neural network.

Our method requires significantly less training data, making it computationally efficient and accessible to experimental groups with limited resources. We validate the approach on both fresh PM6:Y12 and aged PM6:BTP-eC9 organic solar cells, demonstrating its ability to yield meaningful physical insights into device behaviour. These techniques represent a step toward integrating ML into routine materials characterization workflows. By lowering computational and experimental barriers, they open opportunities for rapid feedback during device optimization.

\section{Acknowledgements}
We thank the Deutsche Forschungsgemeinschaft (DFG) for funding this work (Research Unit FOR 5387 POPULAR, project no.\ 461909888).

\bibliographystyle{apsrev4-2}
\bibliography{main}

\clearpage
\onecolumngrid
\begin{center}
\textbf{\Large Supplemental Information}
\end{center}

\setcounter{equation}{0}
\renewcommand{\theequation}{S\arabic{equation}}
\setcounter{figure}{0}
\renewcommand{\thefigure}{S\arabic{figure}}
\setcounter{table}{0}
\renewcommand{\thetable}{S\arabic{table}}
\setcounter{section}{0}
\renewcommand{\thesection}{S\arabic{section}}
\setcounter{page}{1}

\section{Implemented Networks}
\label{si:sec:implemented_networks}
\subsection{Properties Varied}
\begin{table}[H]
\centering
\scalebox{0.7}{
\begin{tabular}{|l|l|c|c|c|c|}
\hline
\textbf{Property}      & \multicolumn{1}{c|}{\textbf{Parameter}}                   & \textbf{Minimum} & \textbf{Maximum} & \textbf{Lin/Log} & \textbf{Units} \\ \hline
Free electron mobility         & Electron mobility y              & $1 \times 10^{-10}$ & $1\times10^{-5}$  & Log & $m^{2}V^{-2}s^{-1}$\\ \hline
Free hole mobility     & Hole mobility z       & $1\times10^{-10}$     & $1\times10^{-5}$      & Log              & $m^{2}V^{-2}s^{-1}$\\ \hline
Shunt resistance       & parasitic/Shunt resistance                                      & $1.42\times10^{3}$    & $28.0$           & Log & $\Omega$             \\ \hline
Series resistance      & parasitic/Series resistance                                     & $1.0$            & $40.0$           & Lin         & $\Omega m^{-2}$     \\ \hline
Electron Urbach energy & Electron tail slope   & $0.03$           & $0.13$           & Lin              & $eV$\\ \hline
Hole Urbach energy     & Hole tail slope       & $0.03$           & $0.13$           & Lin              & $eV$\\ \hline
Electron trap density  & Electron trap density & $1\times10^{18}$      & $1\times10^{26}$      & Log    & $m^{-3}$          \\ \hline
Hole trap density      & Hole trap density     & $1\times10^{18}$      & $1\times10^{26}$      & Log     & $m^{-3}$         \\ \hline
Electron capture cross-section & Free electron to Trapped electron & $1\times10^{-25}$        & $1\times10^{-18}$ & Log  & $m^{2}$ \\ \hline
Electron escape cross-section  & Free electron to Trapped hole     & $1\times10^{-25}$        & $1\times10^{-18}$ & Log  & $m^{2}$\\ \hline
Hole capture cross-section     & Free hole to Trapped electron     & $1\times10^{-25}$        & $1\times10^{-18}$ & Log  & $m^{2}$\\ \hline
Hole escape cross-section      & Trapped hole to Free hole         & $1\times10^{-25}$        & $1\times10^{-18}$ & Log  & $m^{2}$\\ \hline
\end{tabular}}

\caption{Properties of the devices randomly varied in order to create the datasets used to train the Machine Learning Networks. "ActiveLayer" is used as a placeholder as particular to the material used in the active layer.}
\end{table}

\subsection{Point Prediction}
\subsubsection{Hyper-Parameters}

\begin{table}[H]
\centering
\scalebox{0.7}{
\begin{tabular}{|c|c|}
\hline
\textbf{Hyper-Parameter} & \textbf{Value} \\ \hline
Initializer              & he\_normal     \\ \hline
Activation               & SiLU           \\ \hline
Regularization           & None           \\ \hline
Layer Nodes             & 256            \\ \hline
Number of Layers       & 4              \\ \hline
Dropout                  & 0.05           \\ \hline
Epochs                   & 4096           \\ \hline
Patience                 & 128             \\ \hline
Batch Normalisation     & True           \\ \hline
Batch Size              & 1024          \\ \hline
Loss Function           & MSE            \\ \hline
Metrics                  & MAE            \\ \hline
Training Percentage     & 0.8            \\ \hline
Initial earning Rate  & $1\times10^{-4}$    \\ \hline
Decay Rate              & 0.8            \\ \hline
\end{tabular}}
\caption{Hyper-Parameters used to train the point prediction networks}
\end{table}

\subsubsection{Confusion matrices}
\begin{figure}[H]
    \centering
    \includegraphics[width=0.8\linewidth]{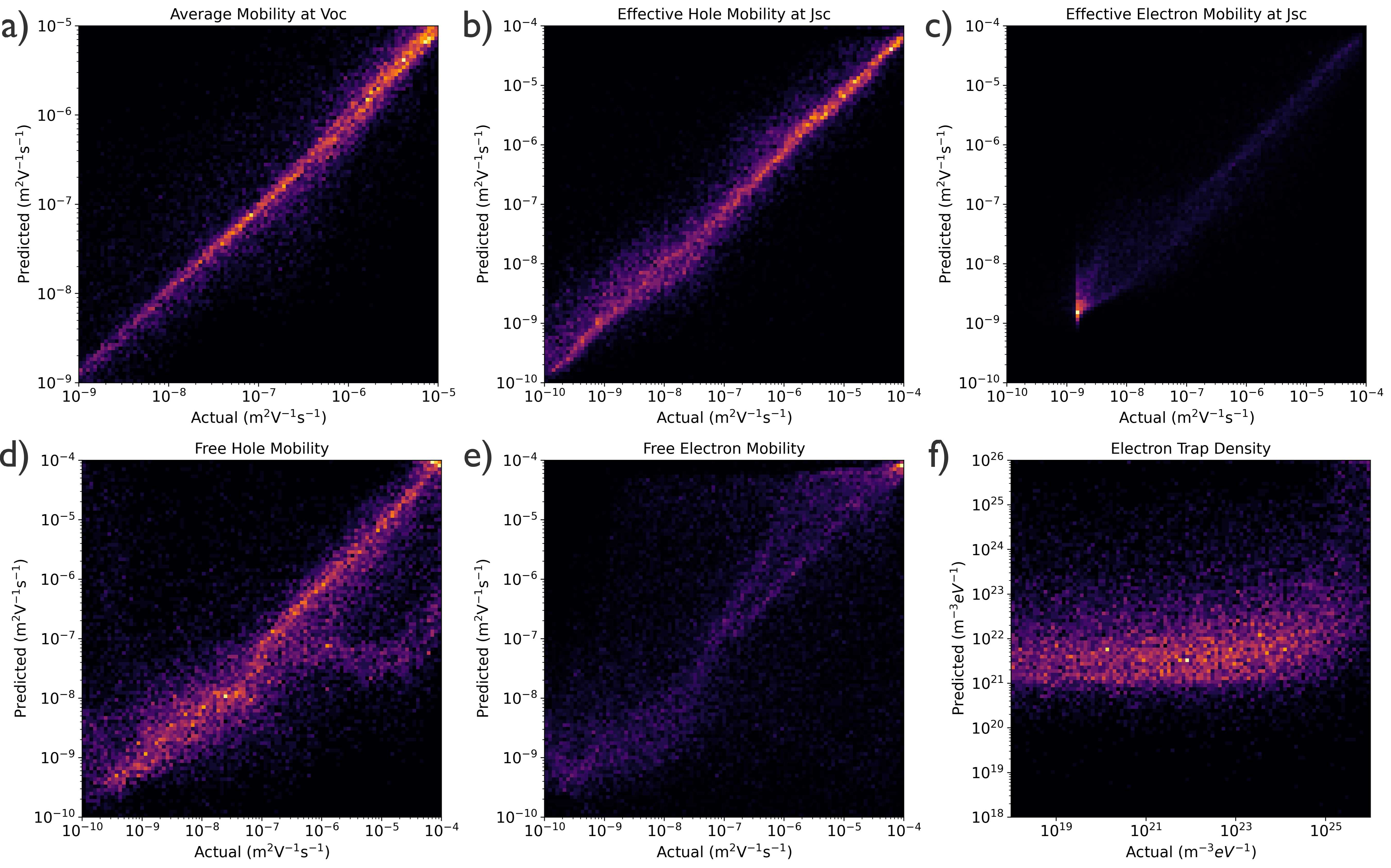}
    \caption{Confusion matrices from the Point method for:
    a) Average mobility at Voc;
    b) Effective Hole mobility at Jsc;
    c) Effective Electron Mobility at Jsc;
    d) Free Hole Mobility;
    e) Free Electron Mobility, and
    f) Electron Trap Density
    }
    \label{fig:enter-label}
\end{figure}

\subsection{Difference Method}
\subsubsection{Hyper-Parameters}

\begin{table}[H]
\centering
\scalebox{0.7}{
\begin{tabular}{|c|c|}
\hline
\textbf{Hyper-Parameter} & \textbf{Value} \\ \hline
Initializer              & he\_normal     \\ \hline
Activation               & SiLU           \\ \hline
Regularization           & None           \\ \hline
Layer Nodes             & 256            \\ \hline
Number of Layers       & 4              \\ \hline
Dropout                  & 0.05           \\ \hline
Epochs                   & 1024           \\ \hline
Patience                 & 16             \\ \hline
Batch Normalisation     & True           \\ \hline
Batch Size              & 16348         \\ \hline
Permutations Limit      & $10\times10^{6}$    \\ \hline
Loss Function           & MSE            \\ \hline
Metrics                  & MAE            \\ \hline
Training Percentage     & 0.8            \\ \hline
Initial earning Rate  & $1\times10^{-5}$    \\ \hline
Decay Rate              & 0.8            \\ \hline
\end{tabular}}
\caption{Hyper-Parameters used to train the difference method networks}
\end{table}

\subsubsection{Confusion Matrices}
\begin{figure}[H]
    \centering
    \includegraphics[width=0.8\linewidth]{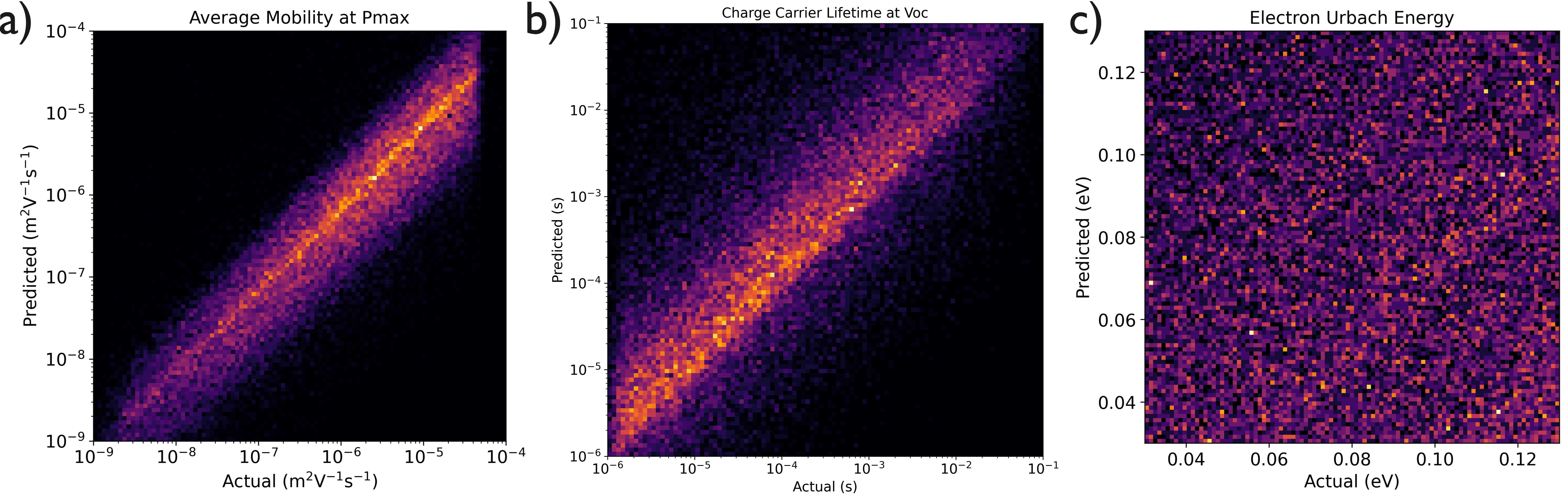}
    \caption{Confusion matrices for;
    a) Average mobility at the maximum power point $\mu_{pmax}$};
    b) Charge carrier lifetime at open circuit voltage $\tau_{voc}$, and
    c) Electron Urbach Energy $E_u^e$.
    \label{fig:Diff_Method_Confusion}
\end{figure}




\subsection{Residual Method}
\subsubsection{Hyper-Parameters}

\begin{table}[H]
\centering
\scalebox{0.7}{
\begin{tabular}{|c|c|}
\hline
\textbf{Hyper-Parameter} & \textbf{Value} \\ \hline
Initializer              & he\_normal     \\ \hline
Activation               & SiLU           \\ \hline
Regularization           & None           \\ \hline
Layer Nodes             & 128            \\ \hline
Number of Layers       & 4              \\ \hline
Number of Residual Blocks & 4           \\ \hline
Dropout                  & 0.05           \\ \hline
Epochs                   & 1024           \\ \hline
Patience                 & 16             \\ \hline
Batch Normalisation     & True           \\ \hline
Batch Size              & 1024          \\ \hline
Permutations Limit      & $10\times10^{6}$    \\ \hline
Loss Function           & MSE            \\ \hline
Metrics                  & MAE            \\ \hline
Training Percentage     & 0.8            \\ \hline
Initial earning Rate  & $8\times10^{-5}$    \\ \hline
Decay Rate              & 0.6            \\ \hline
\end{tabular}}
\caption{Hyper-Parameters used to train the residual method networks}
\end{table}

\subsubsection{Confusion Matrices}
\begin{figure}[H]
    \centering
    \includegraphics[width=0.8\linewidth]{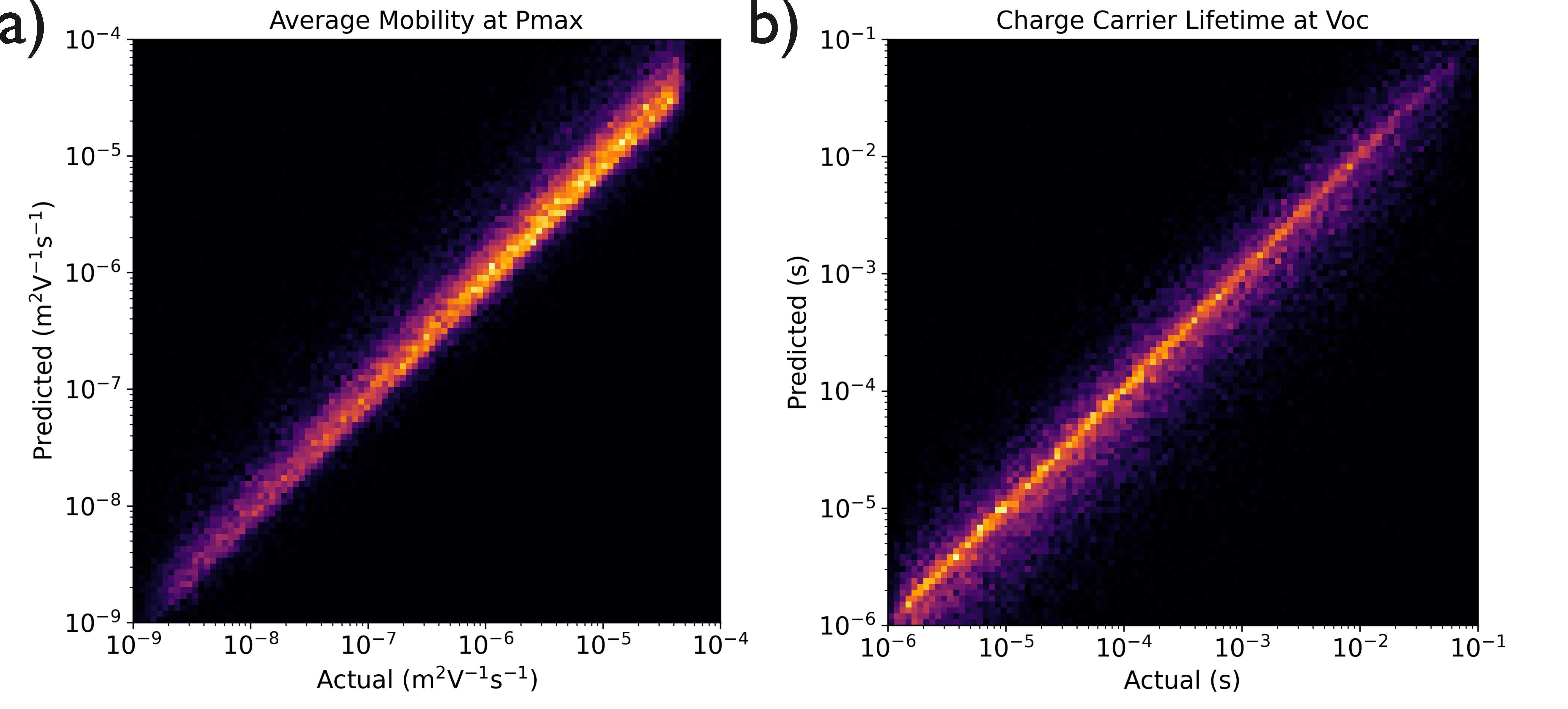}
    \caption{Confusion matrices for;
    a) Average mobility at the maximum power point $\mu_{pmax}$}, and
    b) Charge carrier lifetime at open circuit voltage $\tau_{voc}$
    \label{fig:Res_Method_Confusion}
\end{figure}

\subsection{Network Comparison}
\begin{table}[H]
\centering
\label{tab:Net_Comp}
\begin{tabular}{|l|cc|}
\hline
\multirow{2}{*}{\textbf{Property}}                     & \multicolumn{2}{c|}{\textbf{Root Mean Squared Error}}          \\ \cline{2-3} 
                                                       & \multicolumn{1}{c|}{\textbf{Difference}}  & \textbf{Residual}   \\ \hline
Effective Electron Mobility at Jsc under 1.0 Suns      & \multicolumn{1}{c|}{1.806}               & $1.14\times10^{-3}$ \\ \hline
Effective Hole Mobility at Jsc under 1.0 Suns          & \multicolumn{1}{c|}{0.032}               & $1.02\times10^{-5}$ \\ \hline
Electron Trap Density                                  & \multicolumn{1}{c|}{$2.49\times10^{32}$} & $1.76\times10^{29}$ \\ \hline
Hole Trap Density                                      & \multicolumn{1}{c|}{$1.24\times10^{32}$} & $2.15\times10^{31}$ \\ \hline
Electron Urbach Energy                                 & \multicolumn{1}{c|}{0.060}               & 0.029               \\ \hline
Hole Urbach Energy                                     & \multicolumn{1}{c|}{0.054}               & 0.030               \\ \hline
\end{tabular}
\caption{Root Mean Squared Error (RMSE) values for the difference and residual networks presented within the paper. It can bee seen that overall the residual network achieves improved RMSE when compared to the respective difference network. A difference network has 226,049 trainable parameters, whilst a residual network has 1,007,233 trainable parameters.}
\end{table}

\section{The need to include trap states}
It should be noted that when modelling disordered semiconductors, it is essential to use a model that explicitly accounts for carrier trapping and recombination through trap states. Disordered systems, such as organic semiconductors, exhibit a broad distribution of localized states. These trap states profoundly affect the charge carrier dynamics by capturing carriers, modulating their mobility, and enabling recombination processes that deviate from ideal band-to-band behaviour.

Neglecting trap states in a model can lead to a significant underestimation throughout the device. This is not a minor issue: the electrostatic potential profile within the device is governed by the spatial distribution of charge, and thus errors in carrier density will propagate directly into errors in the calculated internal electric field. This, in turn, affects carrier drift, diffusion, and ultimately the JV curve.

Moreover, key parameters such as the mobility, recombination rate, and charge carrier lifetime are all strongly dependent on the local carrier density. If the model fails to accurately reflect the carrier trapping, any inferred values for these parameters will be biased or unphysical. In particular, mobility can appear artificially low if trapping is neglected. Similarly, recombination rates may be forced to be incorrect to match experimental data.

Therefore, to make meaningful comparisons between simulation and experiment, and to extract physically accurate material parameters, the inclusion of trap-assisted processes is not optional - it is a requirement of any reliable model for disordered semiconductors.

\section{Repositories}

The code utilised, along with documentation for its usage and installation, can be found in the GitHub repository: \href{https://github.com/CaiWilliams/PyOghma_ML}{PyOghma\_ML}.

The datasets used for training the networks presented in this paper are available at the GitHub repository: \href{https://github.com/CaiWilliams/OPV-machine-learning-dataset}{OPV Machine Learning Datasets}

\section{Material Data}
\label{si:sec:md}

\subsection{Device Fabrication}

The patterned ITO substrates were pre-cleaned by ultrasonic baths for 15 minutes in diluted Hellmanex, deionized water, acetone, and 2-propanol sequentially, and then dried with a flowing nitrogen stream. ZnO nanoparticle dispersion (2.5~wt\% in 2-propanol) was spin-coated onto the ITO substrate at 2500~rpm for 60~s, followed by thermal treatment at 200~$^\circ$C for 30~minutes, resulting in a ZnO layer of approximately 37~nm.

The substrates were then transferred into a nitrogen-filled glove box for subsequent active layer and hole transport layer deposition. PM6:BTP-eC9 and PM6:Y12 solutions were prepared in advance, using identical formulations with a 1:1.2 donor-to-acceptor weight ratio. Chloroform was used as the solvent without any additives, with a total concentration of 12~mg/mL. The solutions were stirred at 600~rpm overnight and heated to 45~$^\circ$C prior to deposition.

The warm solutions were spin-coated onto ITO/ZnO substrates at 3000~rpm for 30~s, followed by thermal annealing at 100~$^\circ$C for 10~minutes. The resulting active layer thicknesses were approximately 50--60~nm.

An ethanol-based PEDOT:F solution was used as the hole transport layer. It was spin-coated dynamically at 3000~rpm for 50~s, followed by thermal treatment at 100~$^\circ$C for 5~minutes.

Finally, the devices were completed by thermal evaporation of 150~nm silver (Ag) in a vacuum chamber (vacuum level $\sim 2 \times 10^{-6}$~Torr) using a shadow mask defining an active area of 0.04~cm$^2$.

\subsection{Device Characterisation}
The fabricated devices were characterised using a variety of techniques in order to reveal device properties, such as charge carrier mobility and recombination lifetime. Here are details as to their characterisation.
    
\subsubsection{Current--voltage measurements}
A Keithley 236 SMU was used for voltage application and current measurement. AM1.5G illumination was provided by a Wavelabs LS-2 solar simulator. No aperture was used. The illumination was kept switched on for two seconds per measurement to prevent the sample temperature from increasing. We measured from reverse bias to forward bias with no fixed sweep speed due to enabled autoranging. The measurements were conducted in a nitrogen-filled glovebox.
    
\subsubsection{IMVS and IMPS}

Modulated and continuous illumination was provided by an Omicron A350 diode laser with a center wavelength of 515 nm. A Zurich Instruments MFLI lock-in amplifier with MF-IA, MF-MD, and MF-5FM options was used to measure sample current and voltage as well as to provide voltage to modulate the laser. The illumination intensity was varied using neutral-density filters mounted in a Thorlabs motorized filter wheel FW102C combined with a continuously variable neutral-density filter wheel. For IMPS and IMVS measurements, the amplitude of modulated illumination was chosen to be $10\%$ of the bias illumination intensity to ensure small perturbation conditions. Light intensity was continuously monitored using a Newport 818-BB-21 silicon photodetector. 
To ensure open-circuit conditions during IMVS measurements, a 10~M\textOmega~input resistance was used, while for IMPS, the measurements were performed close to short circuit. There, the current was measured over the input resistance of 50~\textOmega. 

\begin{figure}[H]
    \centering
    \includegraphics[width=0.5\linewidth]{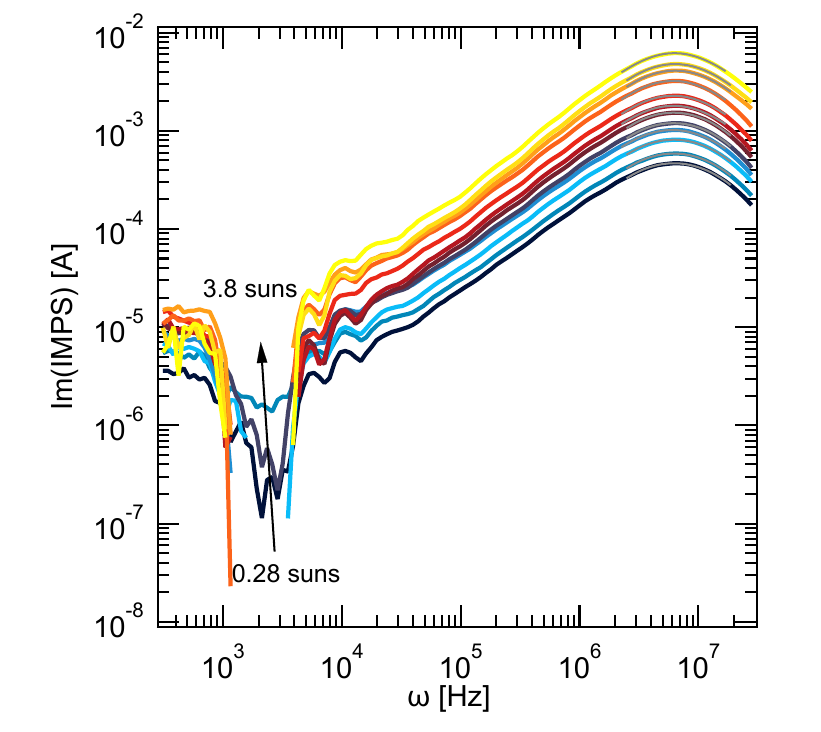}
    \caption{Imaginary part of experimental IMPS data with Gaussian fits indicated by grey lines at each respective peak.}
    \label{fig:Si-imps}
\end{figure}

\begin{figure}[H]
    \centering
    \includegraphics[width=0.5\linewidth]{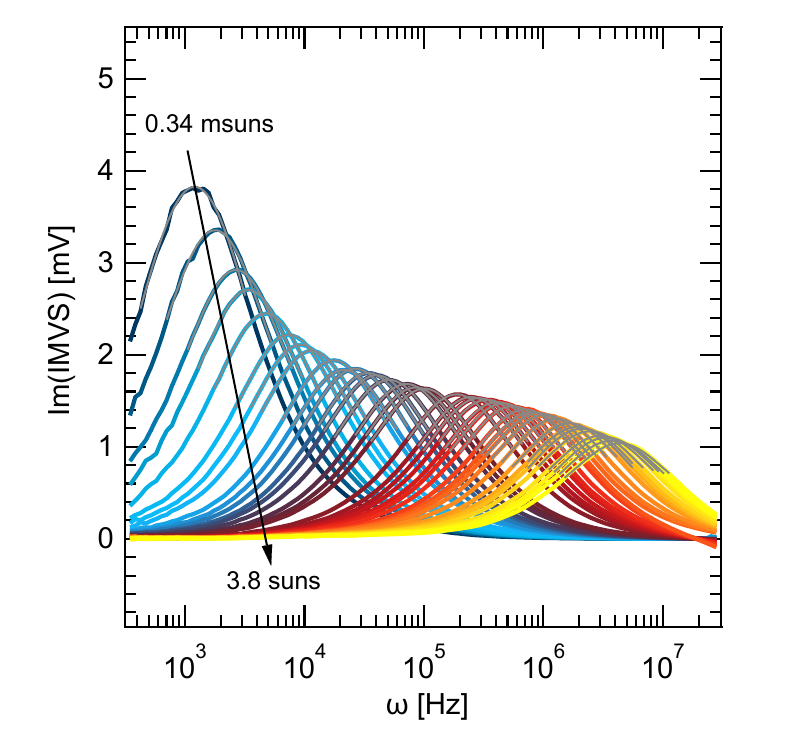}
    \caption{Imaginary part of experimental IMVS data with Gaussian fits indicated by grey lines at each respective peak.}
    \label{fig:si-imvs}
\end{figure}

The corresponding data are shown in Figures~\ref{fig:Si-imps} and \ref{fig:si-imvs}. The characteristic frequency $\omega_\mathbf{c}$ -- for both IMVS and IMPS -- was determined from the peak of the imaginary component and was used to evaluate the recombination lifetime $\tau_\mathrm{rec}$ (for IMVS) and charge carrier mobility $\mu$ (for IMPS). The recombination lifetime was evaluated according to:\cite{set2015analytical}
\begin{equation}\begin{split}\label{eq:tau_IMVS}
    \frac{\mathrm{d} \ln\omega_\mathbf{c}\left(\Phi\right)}{\mathrm{d} \ln\Phi} &= 1-\frac{1}{\delta} , \\
    \tau_\mathrm{rec} &= \frac{\delta}{\omega_\mathbf{c}} . \\
\end{split}\end{equation}
Here, the recombination order $\delta$ was determined using the light intensity-dependence of the characteristic frequency $\omega_\mathbf{c}\left(\Phi\right)$.

For IMPS, the mobility $\mu$ was evaluated according to:\cite{nojima2019modulated} 
\begin{align}\label{eq:mu_IMPS}
    \mu=\frac{\omega_{c}\,L^2}{2\,V_\mathrm{bi}} . 
\end{align}
Here, $L$ is the thickness of the photoactive layer, and $V_\mathrm{bi}$ is the build-in voltage, estimated as 1.15~V.

\subsubsection{SCLC} 
\begin{figure}[H]
    \centering
    \includegraphics[width=0.5\linewidth]{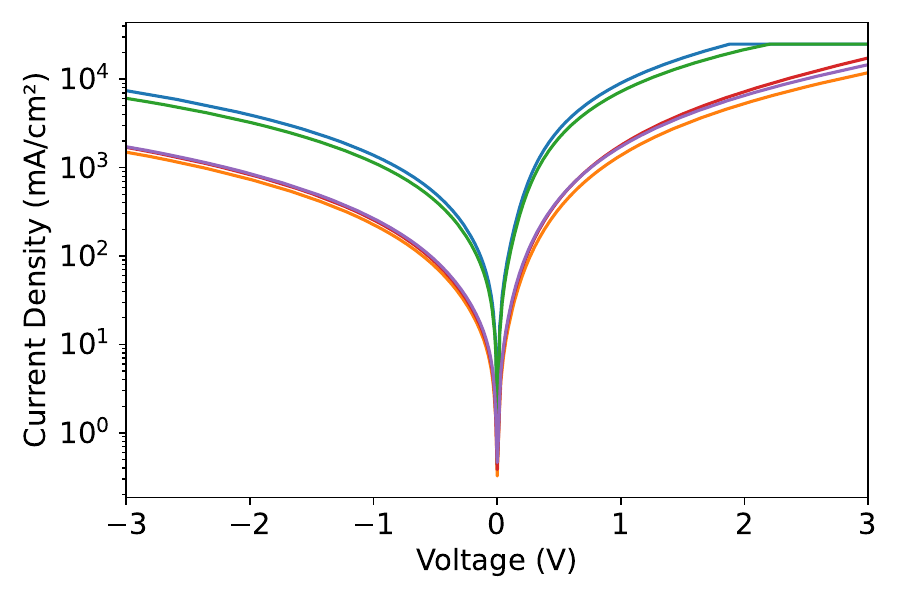}
    \caption{PM6:Y12 blend-based electron only device, different colours represent different pixels of the device.}
    \label{fig:SI-SCLC-e}
\end{figure}

\begin{figure}[H]
    \centering
    \includegraphics[width=0.5\linewidth]{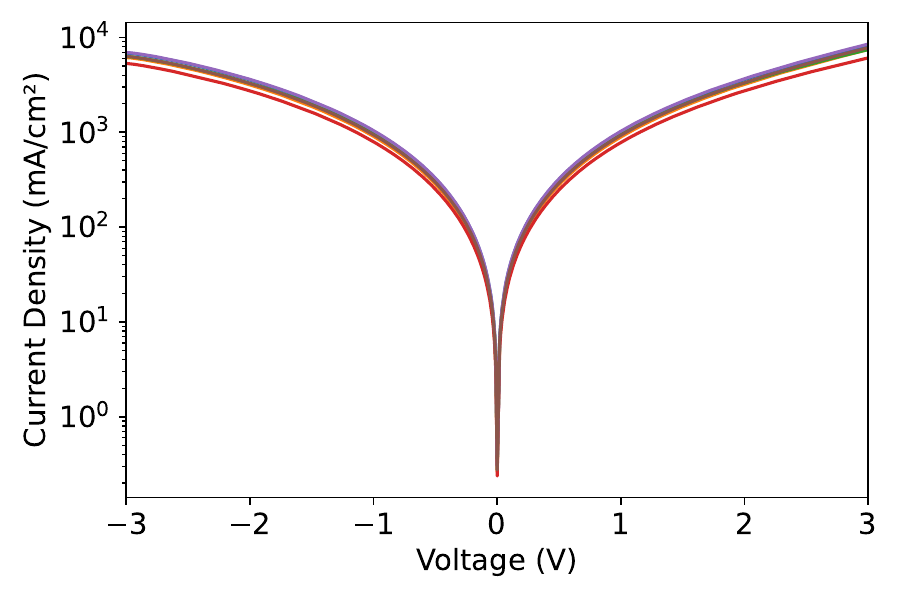}
    \caption{PM6:Y12 blend-based hole only device, different colours represent different pixels of the device.}
    \label{fig:SI-SCLC-h}
\end{figure}

The hole- and electron-only devices were fabricated to determine the charge carrier mobility. The configuration of ITO/PEDOT:PSS/Active layer/PEDOT:F/Ag and ITO/ZnO/Active layer/PDINO/Al was used for the hole-only and electron-only devices, respectively. The current was measured using a Keithley 236 SMU, which was also used to apply voltage to the samples. The corresponding data are shown in Figures~\ref{fig:SI-SCLC-e} and \ref{fig:SI-SCLC-h}. The mobilities were determined using the Mott-Gurney equation:\cite{mott1948electronic}
    \begin{equation}
        J=\frac{9}{8}\epsilon_{r}\epsilon_{0}\mu_{e,h}\frac{V^2}{L^3} , 
    \end{equation}
where $J$ is the current density, $\epsilon_{r}$ is the relative permittivity, assumed to be 3.5, $\epsilon_{0}$ is the vacuum permittivity, $\mu_{e,h}$ is the mobility of electrons (e) or holes (h), $V$ is the applied voltage, and $L$ is the active layer thickness. 

\subsubsection{Refractive Index and Extinction Coefficient}

The optical constants n and k of ZnO, PEDOT:F, as well as PM6:Y12 and PM6:BTP-eC9 blend films were determined using variable-angle spectroscopic ellipsometry (VASE) measurements of individual layers deposited on glass substrates. The backsides of these samples were roughened before measurements to suppress the effect of backside reflections. The VASE measurements were performed with an M2000 T-Solar ellipsometer by J.\ A.\ Woollam at 7 different angles between 45° - 75° in steps of 5° and in a spectral range from approximately 0.7 - 5.1~eV, as provided by a Xe-arc lamp. The measurement area is between 200 – 300 \textmu m depending on the angle of incidence. Three arbitrary but well-separated spots were measured on each sample to be correlated in the processing of the measurement data. The values of the optical constants of the thin films were extracted from a fitted optical model of the dielectric functions using the CompleteEASE Software by J.\ A.\ Woollam. To that end, a reference glass substrate was measured as well. 

The modelling approach for all materials involved an initial approximation of the spectral range below the main absorption onset using a Cauchy function, as estimated from the line shape of the measured data. For the thin films, this also yielded an initial guess of the film thickness and surface roughness. Thereafter, absorption was first approximated via the reparametrisation of the imaginary part of the dielectric function using a B-Spline function from which the real part of the dielectric function was calculated via a Kramers--Kronig transformation. The resulting line shapes of the dielectric function were then parametrised using oscillator models. For the main absorption onsets, a Cody--Lorentz oscillator function was used. All other absorption features were fitted with Gaussian oscillators.

\begin{figure}[H]
    \centering
    \includegraphics[width=0.5\linewidth]{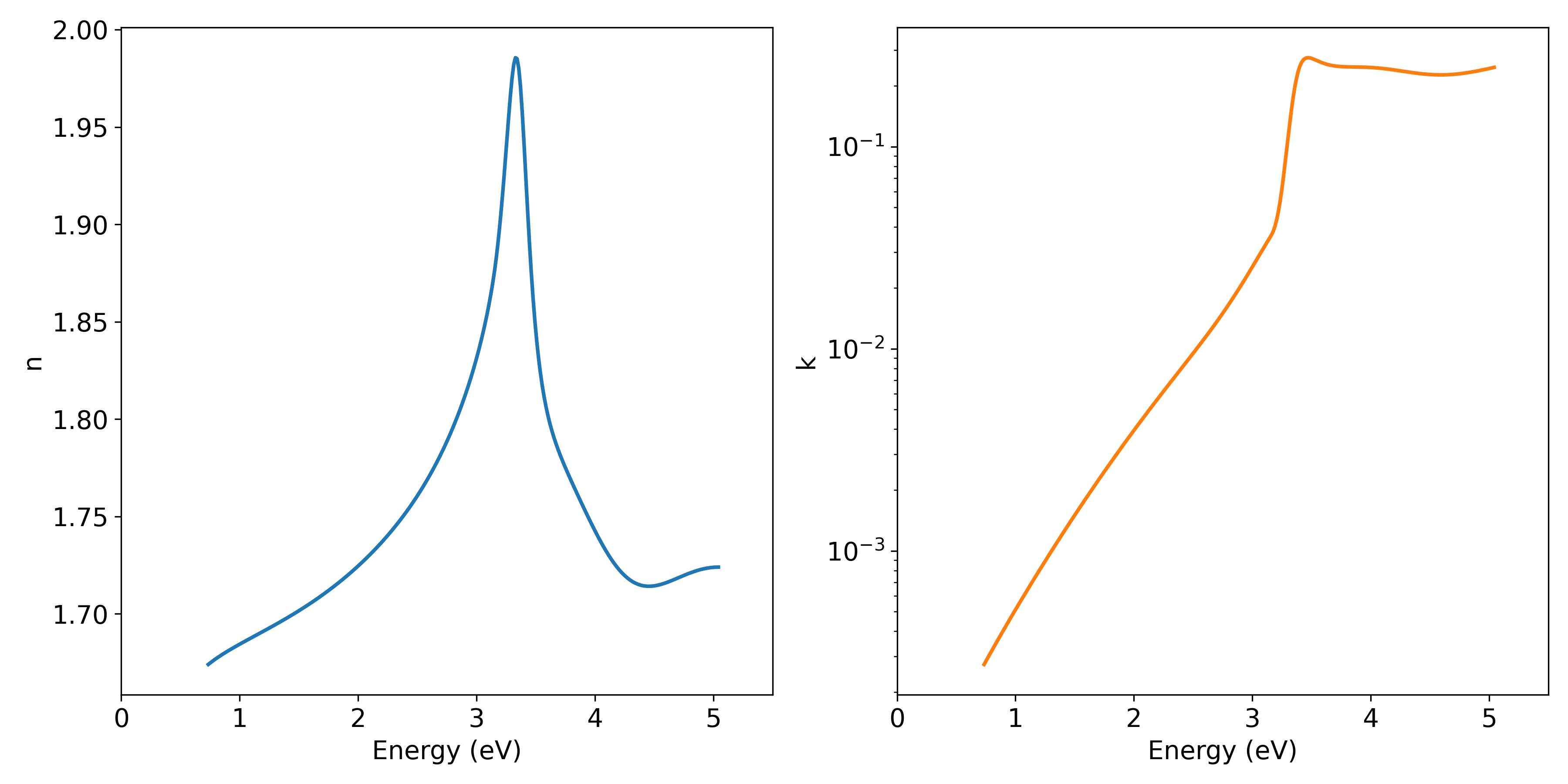}
    \caption{Refractive index (n) and extinction coefficient (k) data for a ZnO thin film.}
    \label{fig:si:nk:zno}
\end{figure}

\begin{figure}[H]
    \centering
    \includegraphics[width=0.5\linewidth]{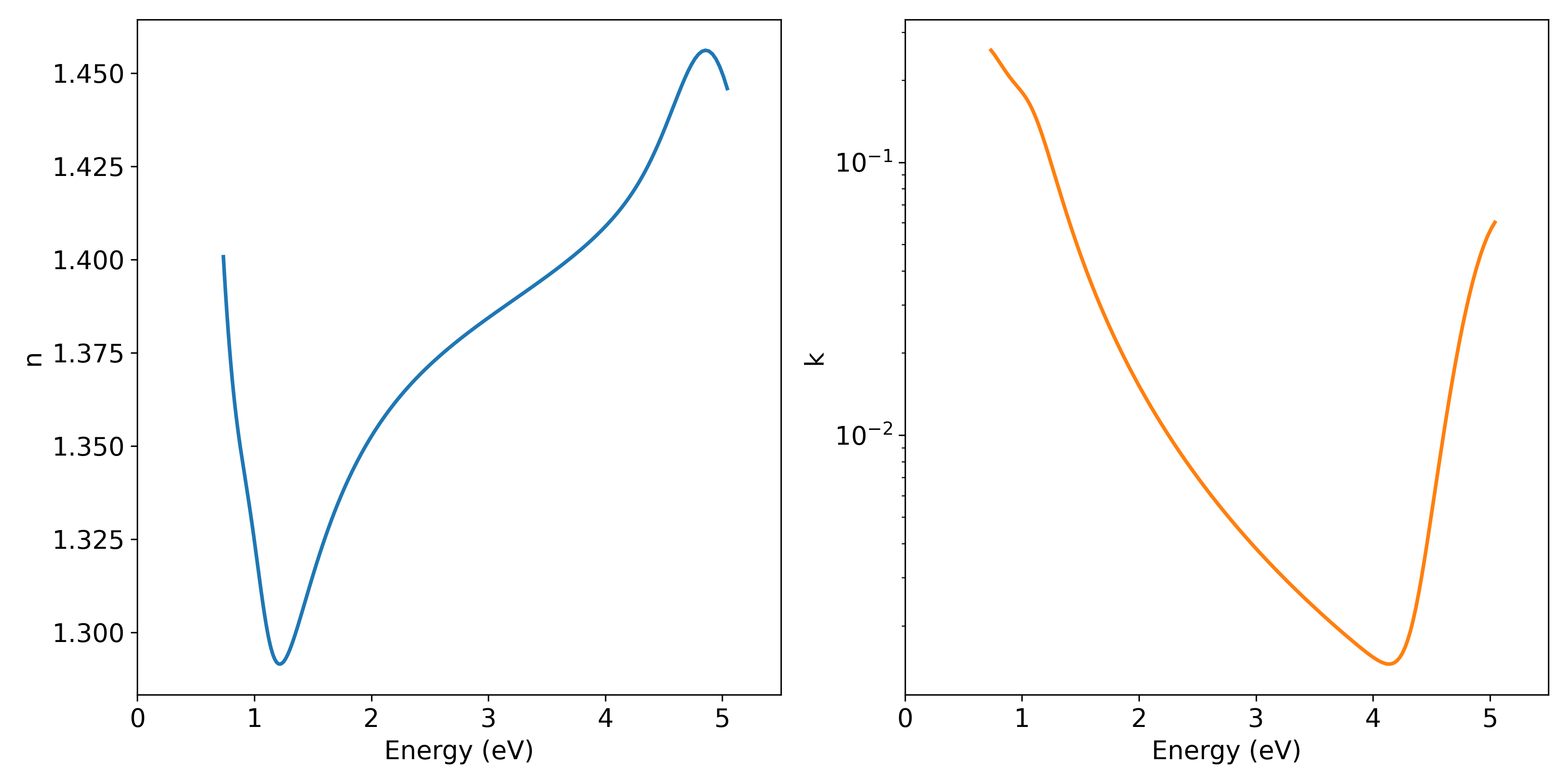}
    \caption{Refractive index (n) and extinction coefficient (k) data for a PEDOT:F thin film.}
    \label{fig:si:nk:pedotf}
\end{figure}

\begin{figure}[H]
    \centering
    \includegraphics[width=0.5\linewidth]{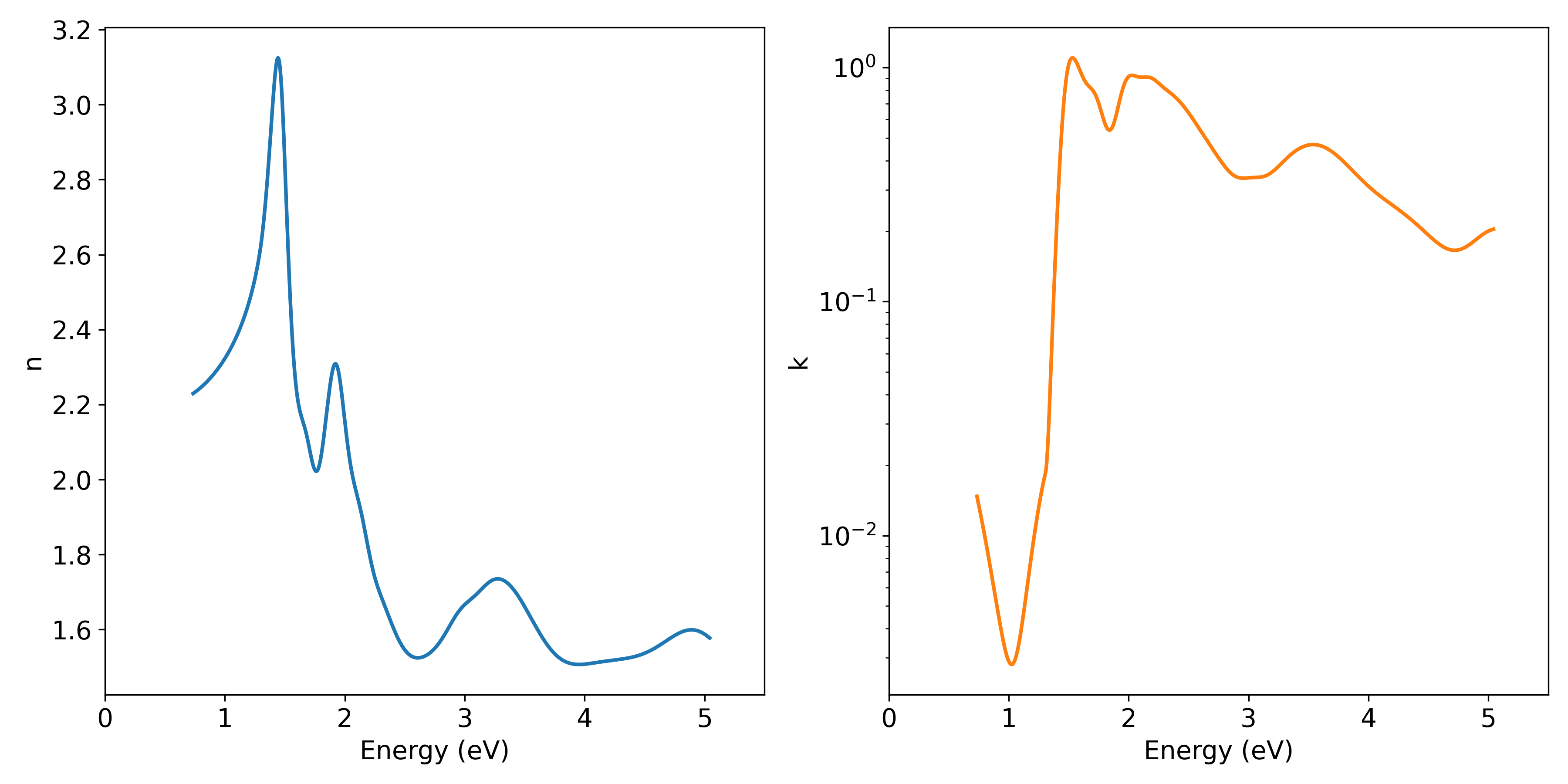}
    \caption{Refractive index (n) and extinction coefficient (k) data for a PM6:BTP-eC9 thin film.}
    \label{fig:si:nk:pm6btpec9}
\end{figure}

\begin{figure}[H]
    \centering
    \includegraphics[width=0.5\linewidth]{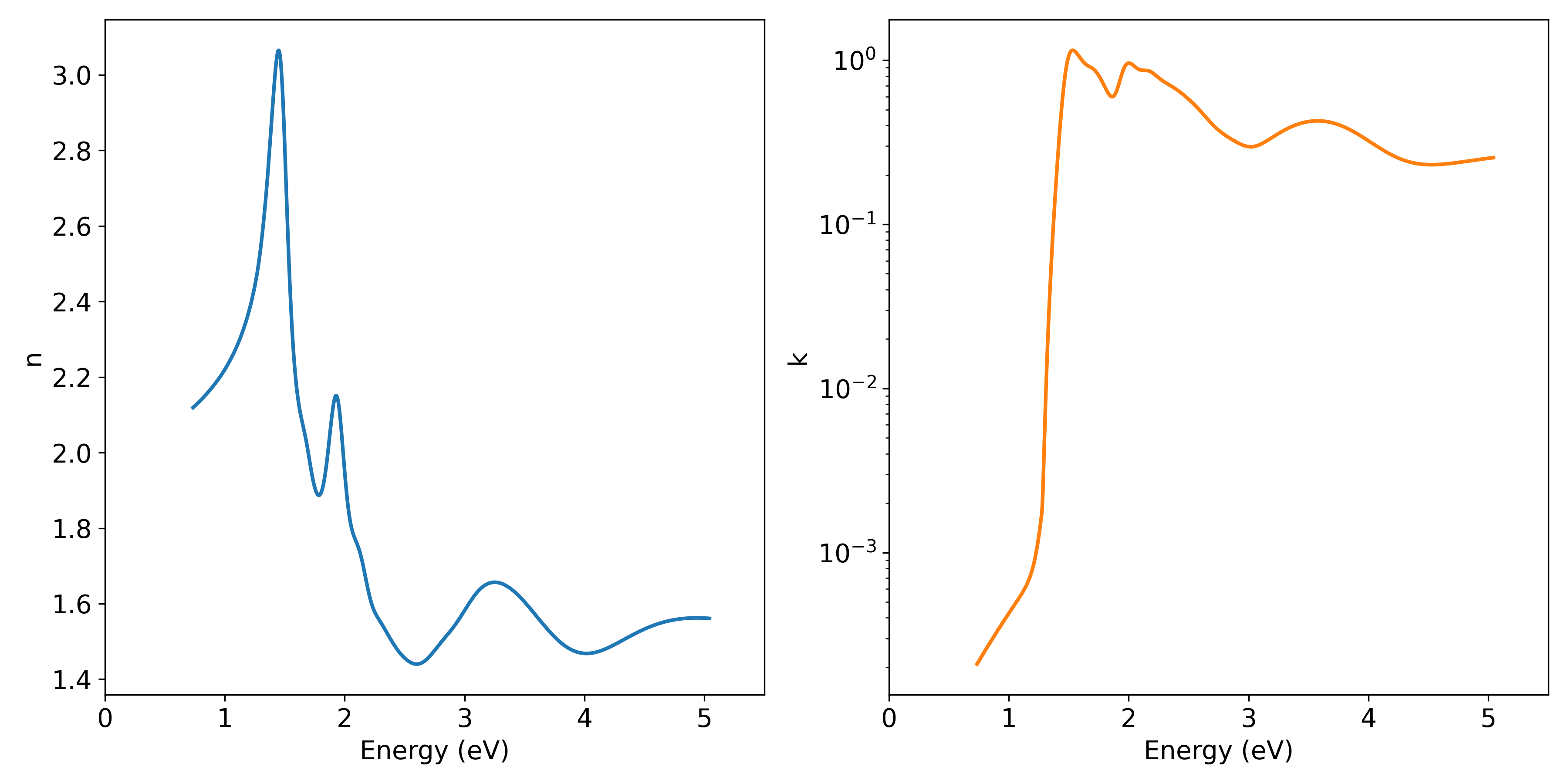}
    \caption{Refractive index (n) and extinction coefficient (k) data for a PM6:Y12 thin film.}
    \label{fig:si:nk:pm6y12}
\end{figure}

\end{document}